# EFFECTS OF BULK COMPOSITION ON THE ATMOSPHERIC DYNAMICS ON CLOSE-IN EXOPLANETS

Xi Zhang[1] and Adam P. Showman[2,3]

[1]Department of Earth and Planetary Sciences, University of California Santa Cruz, CA 95064, USA xiz@ucsc.edu
[2]Department of Atmospheric and Oceanic Sciences, School of Physics, Peking University, Beijing, 100871, China and
[3]Department of Planetary Sciences and Lunar and Planetary Laboratory, University of Arizona, AZ 85721, USA

## Abstract

Super Earths and mini Neptunes likely have a wide range of atmospheric compositions, ranging from low-molecular mass atmospheres of $H_2$ to higher molecular atmospheres of water, $CO_2$, $N_2$, or other species. Here, we systematically investigate the effects of atmospheric bulk compositions on temperature and wind distributions for tidally locked sub-Jupiter-sized planets, using an idealized 3D general circulation model (GCM). The bulk composition effects are characterized in the framework of two independent variables: molecular weight and molar heat capacity. The effect of molecular weight dominates. As the molecular weight increases, the atmosphere tends to have a larger day-night temperature contrast, a smaller eastward phase shift in the thermal phase curve and a smaller zonal wind speed. The width of the equatorial super-rotating jet also becomes narrower and the "jet core" region, where the zonal-mean jet speed maximizes, moves to a greater pressure level. The zonal-mean zonal wind is more prone to exhibit a latitudinally alternating pattern in a higher-molecular-weight atmosphere. We also present analytical theories that quantitatively explain the above trends and shed light on the underlying dynamical mechanisms. Those trends might be used to indirectly determine the atmospheric compositions on tidally locked sub-Jupiter-sized planets. The effects of the molar heat capacity are generally small. But if the vertical temperature profile is close to adiabatic, molar heat capacity will play a significant role in controlling the transition from a divergent flow in the upper atmosphere to a jet-dominated flow in the lower atmosphere.

*Subject headings:* atmospheric effects – hydrodynamics – planets and satellites: atmospheres – planets and satellites: general – methods: analytical – methods: numerical

## 1. INTRODUCTION

According to recent *Kepler* statistics, most of the population of known transiting planets (including planetary candidates) around GFKM stars is dominated by planets of size between one and four Earth radii (Howard et al. 2012; Fressin et al. 2013; Petigura et al. 2013). From small to large, those planets can be basically categorized into Earth-size planets, super Earths and mini Neptunes (Fressin et al. 2013). Many of them are expected to harbor atmospheres, several of which have been characterized, including GJ1214b (Kreidberg et al. 2014), GJ436b (Knutson et al. 2014a), HD 97658b (Knutson et al. 2014b), GJ 3470b (Ehrenreich et al. 2014), and HAT-P-11b (Fraine et al. 2014). When the planet is smaller than Jupiter and especially Neptune or Uranus, the atmosphere might no longer be solely dominated by hydrogen (e.g., Moses et al. 2013). Compositional diversity of the atmospheres emerges, as seen on the terrestrial planets in the Solar System. However, identifying the compositions of those atmospheres is difficult because the transit spectroscopic measurements are influenced by hazes or clouds. In fact, four out of the above five planets show featureless transmission spectra (Kreidberg et al. 2014; Knutson et al. 2014a; Knutson et al. 2014b; Ehrenreich et al. 2014), except HAT-P-11b that exhibits water absorption in a hydrogen environment (Fraine et al. 2014).

Theoretically, the observable outer envelopes of the atmospheres of the sub-Jupiter-size planets could exhibit a variety of compositions (e.g., Pierrehumbert 2013; Moses et al. 2013; Lissauer et al. 2014; Hu & Seager 2014). The primordial constituents of the atmosphere are mostly determined by the metallicity of the protoplanetary disk (e.g., Marboeuf et al. 2008; Johnson et al. 2012), the details of gas accretion during planetary formation, location of the planetary formation, and subsequent migration tracks (e.g., Rogers et al. 2011). The atmospheric bulk compositions will also be greatly influenced by many processes during evolution after planetary formation. Those processes include planetary and asteroid impacts, volcanism and outgassing (Kite et al. 2009), plate tectonics (Elkins-Tanton & Seager 2008), atmosphere-surface interaction, atmosphere-ocean exchange, atmospheric escape (Lopez et al. 2012; Owen & Wu 2013; Hu et al. 2015), and possible biospheric processes (Domagal-Goldman et al. 2011; Seager et al. 2013). Atmospheric dynamics and chemistry will also play significant roles in determining the observable atmospheric composition (e.g., Liang et al. 2004; Cooper & Showman 2005; Segura et al. 2007; Zahnle et al. 2009; Line et al. 2010; Moses et al. 2011; Kempton et al. 2012; Moses et al. 2013; Parmentier et al. 2013; Hu & Seager 2014).

As a result, we expect to observe a large compositional diversity among sub-Jupiter-size exoplanets. There could be hydrogen-dominated atmospheres as on Jupiter and hot Jupiters; more metal-rich (but still hydrogen-dominated) atmospheres like Uranus and Neptune; evaporated atmospheres dominated by helium (proposed for GJ436b, Hu et al. 2015); or carbon dioxide atmosphere like those of Venus and Mars; water worlds (proposed for GJ1214b, Rogers & Seager 2010; Miller-Ricci & Fortney 2010; Nettelmann et al. 2011); nitrogen and oxy-gen atmosphere like Earth; nitrogen and methane atmosphere like Saturn's moon Titan; atmosphere of sulfur compounds like Jupiter's moon Io, carbon monoxide or hydrocarbon atmospheres (Hu & Seager 2014), or some metallic-based or silicon-based atmospheres if the planet is very close to the host star (Schaefer & Fegley 2009; Miguel et al. 2011; Schaefer et al. 2012). More quantitatively, given the metallicity and effective temperature, thermochemical and photochemical calculations have predicted a variety of possible bulk constituents for exo-planetary atmospheres (e.g., Moses et al. 2013; Hu & Seager 2014).

Most known exoplanets are close to their host stars and expected to be tidally locked with a synchronized rotation around the central star. Weather on those planets is intriguing. Previous studies mostly focused on the hydrogen atmospheres of hot Jupiters (e.g., Cooper & Showman 2005; Dobbs-Dixon & Lin 2008; Showman et al. 2009; Rauscher & Menou 2010; Heng et al. 2011; Perna et al. 2010; Mayne et al. 2014), and Earth-like environments for Earth-size planets in the habitable zone (e.g., Merlis & Schneider 2010; Hu & Yang 2014; Yang et al. 2014; Wordsworth 2015). Lewis et al. (2010) has considered the effect of different atmospheric opacity sources on the general circulation for GJ436b, but the bulk composition of the atmosphere is still hydrogen. For another super Earth GJ1214b, whose bulk atmospheric composition is still under debate (Rogers & Seager 2010; Miller-Ricci & Fortney 2010; Nettelmann et al. 2011), Menou (2011) simulated the general circulation patterns and thermal phase curves for the hydrogen atmospheres of various metallicity, as well as for water-dominated atmospheres. He pointed out that the molecular weight and opacity have strong effects on the weather pattern. Zalucha et al. (2013) investigated the behavior this planet might exhibit if its atmosphere were water vapor. With a more realistic radiative transfer scheme, the dynamical simulations by Kataria et al. (2014) further explored the effects of metallicity in hydrogen atmospheres as well as possible compositions including water, carbon dioxide and their mixtures with hydrogen. They found that atmospheres with a lower mean-molecular weight have smaller day-night and equator-to-polar temperature contrast than higher mean-molecular weight atmospheres. They also showed that a water atmosphere exhibits an equatorial prograde jet, but a carbon-dioxide atmosphere is dominated by two high-latitude jets, suggesting significant bulk composition effects on the atmospheric dynamics. Charnay et al. (2015a) investigated the impact of different metallicities in a hydrogen atmosphere on vertical mixing and circulation pattern on GJ1214b. Charnay et al. (2015b) included the cloud tracers and its radiative feedbacks in the dynamical model but the atmosphere is still primarily dominated by hydrogen. To date, there has not been a systematic study on all possible bulk compositions, ranging from lighter molecules such as $H_2$ and He to heavier molecules such as $N_2$ and $CO_2$, to investigate the underlying mechanism of their effects on the atmospheric dynamics and general circulation of tidally locked planets.

There are three important properties of an atmospheric species that could influence the atmospheric dynamics: opacity, molecular weight, and heat capacity. Opacity fundamentally controls the heating and cooling rates in the atmosphere, which drives the dynamics. Molecular weight greatly affects the scale height of an atmosphere, the buoyancy frequency, atmospheric wave properties such as gravity wave speed, and the deformation radius, which is a typical dynamical length scale in the atmosphere. Heat capacity also affects the radiative heating and cooling rates, buoyancy, wave speed and deformation radius, but in a different way from that of the molecular weight. Given the complicated roles the bulk composition could play, it is difficult to disentangle the underlying mechanisms if we consider all those effects together. Instead, here we use an idealized general circulation model (GCM) to study individual effects. Specifically, we investigate the effects of the molecular weight and heat capacity while holding fixed the day-night heating scheme. Note that gas opacity is commonly dominated by trace species that do not significantly affect the molecular mass, e.g., water and carbon dioxide in the case of Earth; water, carbon monoxide, and methane in the case of hot Jupiters. Jupiter itself has significant opacity contribution from the background hydrogen, but also from trace hydrocarbons including methane, ethane, and acetylene (Zhang et al. 2013). Therefore, we leave a thorough investigation of the opacity effect for the future.

This study has important implications for observations. We aim to build quantitative theories for the bulk composition effects using a series of analytical expressions and predict some observable signatures, including the spatial maps of wind and temperature, day-night temperature differences and thermal phase curves. It is expected that the atmospheric dynamics will have significant effects on these observables. For tidally locked planets, the day-night temperature difference and the shape of the thermal phase curve can be greatly influenced by the day-night heat redistribution efficiency and the presence of an equatorial prograde jet (e.g., Showman & Guillot 2002; Knutson et al. 2007; Cowan & Agol 2011; Perez-Becker & Showman 2013; Komacek & Showman 2016). The atmospheric wind speed might be directly measured through ultra-high resolution Doppler mapping techniques that have been applied to hot Jupiter HD209458b (Snellen et al. 2010) and HD189733b (Wyttenbach et al. 2015; Louden & Wheatley 2015). On the other hand, measuring phase curves and winds might provide some hints on the properties of the bulk composition in the atmosphere. This may serve as an alternative way to overcome the current difficulties in determining the atmospheric composition of transiting super Earths/mini Neptunes that exhibit featureless spectra, such as GJ1214b.

This paper is organized as follows. In Section 2, we will first provide some theoretical background on the effects of compositions on atmospheric dynamics. We will introduce the atmospheric circulation model and qualitatively summarize the main results from our simulations in Section 3. Then we will dig into the underlying physical mechanisms by which the temperature and wind patterns are affected by molecular weight (Section 4) and molar heat capacity (Section 5). We will also derive a series of analytical expressions to interpret the simulation results. This paper ends with a conclusion and implications for the observations in Section 6. The analytical treatment of temperature, wind and thermal phase shift are detailed in the Appendices.

Molecular weight ($\mu$, g mol$^{-1}$) and molar heat capacity at constant pressure ($c_p$, J mol$^{-1}$K$^{-1}$) for typical atmospheric bulk compositions. Thermodynamical data are from Chase (1998).

| Composition | Molecular Weight ($\mu$) | $c_p$ (300 K) | $c_p$ (600 K) | $c_p$ (1000 K) | $c_p$ (2000 K) |
|---|---|---|---|---|---|
| **H$_2$** | 2.02 | 28.85 | 29.43 | 30.21 | 34.28 |
| **He** | 4.00 | 20.79 | 20.79 | 20.79 | 20.79 |
| **CH$_4$** | 16.04 | 35.71 | 51.17 | 71.80 | 94.40 |
| **H$_2$O** | 18.02 | 33.60 | 36.88 | 41.27 | 51.18 |
| **CO** | 28.01 | 29.14 | 30.87 | 33.18 | 36.25 |
| **N$_2$** | 28.01 | 29.13 | 30.66 | 32.70 | 35.97 |
| **O$_2$** | 32.00 | 29.39 | 31.74 | 34.87 | 37.74 |
| **CO$_2$** | 44.01 | 37.22 | 44.54 | 54.31 | 60.35 |

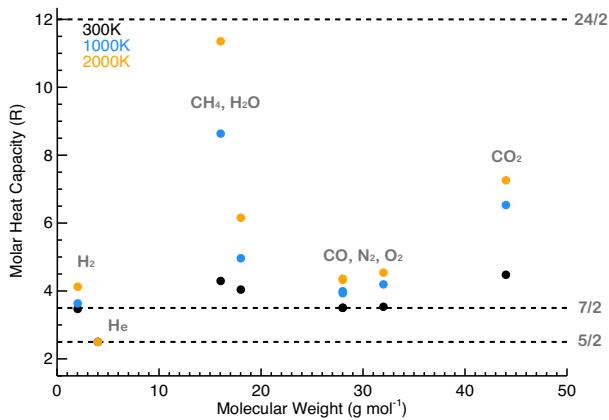

Fig. 1.— Diagram of Molecular weight and molar heat capacity for typical bulk constituents of atmospheres for super Earths and mini Neptunes.

## 2. THEORETICAL BACKGROUND

### 2.1. *Diversity of the Atmospheric Composition*

Thermochemical and photochemical models (e.g., Moses et al. 2013; Hu & Seager 2014) predict a variety of possible bulk constituents, depending on the metallically of the atmosphere, carbon to oxygen ratio, stellar flux, and atmospheric evolution. Plausible constituents exhibit a wide range of mean-molecular weight ($\mu$) and molar heat capacity at constant pressure ($c_p$), summarized in Table 1 and Fig. 1. Molecular weight can vary by a factor of $\sim$20, from 2 g mol$^{-1}$ to 44 g mol$^{-1}$. The heat capacity, however, can only vary by a factor of $\sim$4, from $5R/2$ for helium to about $11R$ for high-temperature methane, where $R = 8.314$ g mol$^{-1}$ K$^{-1}$ is the molar gas constant.

The molecular weight and molar heat capacity are independent properties of a gas molecule. The latter is solely controlled by the degrees of freedom of the molecule. The monatomic gas, helium, which could conceivably comprise the background atmospheric gas as a result of atmospheric escape processes (Hu et al. 2015), has the lowest $c_p$ ($5R/2$). Diatomic gases, such as hydrogen, carbon monoxide, nitrogen and oxygen, have molar heat capacities ranging from $7R/2$ at low temperature up to $9R/2$ at high temperature for the vibrational modes to be fully excited. Multi-atomic molecules, such as a triatomic linear molecule carbon dioxide and more complex-structured molecules like water and methane, usually contain a larger number of degrees of freedom and therefore span a large range of $c_p$. For example, the molar heat capacity of methane can reach more than $11R$ at 2000 K, although thermochemical equilibrium prefers forming carbon monoxide instead of methane at high temperature (e.g., Moses et al. 2013). From thermochemical calculations, CO and CH$_4$ will never dominate the bulk atmosphere, but their contribution to the mean molecular weight and molar heat capacity could be significant.

### 2.2. *Effects on the Atmospheric Dynamics*

Through what physical mechanisms could the molecular weight and molar heat capacity influence the atmospheric dynamics on a tidally locked planet, including observational signatures such as temperature and wind speed? For a tidally locked planet in the moderate temperature range, if the magnetohydrodynamics (MHD) effect can be ignored, the photospheric dynamics is approximately governed by the hydrostatic primitive equations. In pressure ($p$) coordinate, the equations are:

$$\frac{D\vec{\mathbf{u}}}{Dt} + f\hat{\mathbf{k}} \times \vec{\mathbf{u}} + \nabla_p \Phi = F_d \qquad (1a)$$

$$\frac{\partial \Phi}{\partial p} = -\frac{1}{\rho} \qquad (1b)$$

$$\nabla_p \cdot \vec{\mathbf{u}} + \frac{\partial \omega}{\partial p} = 0 \qquad (1c)$$

$$\frac{DT}{Dt} - \frac{\omega \mu}{\rho c_p} = \frac{q\mu}{c_p} \qquad (1d)$$

$$p = \frac{\rho R T}{\mu} \qquad (1e)$$

where $\rho$ is density; $\mu$ is molecular weight; $T$ is temperature; $c_p$ is molar heat capacity in units of J mol$^{-1}$ K$^{-1}$; $\omega = Dp/Dt$ is vertical velocity; $\Phi = gz$ is geopotential where $g$ is gravitational acceleration and $z$ is altitude; $f = 2\Omega \sin\phi$ is Coriolis parameter where $\Omega$ is the planetary rotation rate and $\phi$ is latitude; $\nabla_p$ is the horizontal gradient at constant pressure; $\vec{\mathbf{u}} = (u, v)$ is the horizontal velocity at constant pressure where $u$ is the zonal (east-west) velocity and $v$ is the meridional (north-south) velocity; $q$ is the radiative heating/cooling rate in units of J kg$^{-1}$ s$^{-1}$. $D/Dt = \partial/\partial t + \vec{\mathbf{u}} \cdot \nabla_p + \omega \partial/\partial p$ is the material derivative; $F_d$ is a drag term representing missing physics such as sub-grid turbulent mixing (e.g., Li & Goodman 2010; Youdin & Mitchell 2010). Here we have assumed that the atmosphere is in hydrostatic balance. This is generally a good assumption (Showman et al. 2008; Showman et al. 2008). For hot Jupiters, the non-hydrostatic models from Dobbs-Dixon & Agol (2013) and Mayne et al. (2014) exhibit qualitatively similar behavior (e.g., overall structure of the superrotating jet) to the hydrostatic models.

Both the molecular weight and molar heat capacity appear in the thermodynamic equation (1d) and the molecular weight is also present the equation of state (1e). A change of the thermodynamic properties in the atmosphere (e.g., temperature) will directly feed back to the geopotential field and lead to a change of atmospheric motion. The major effects on the atmospheric dynamics can be categorized into radiative effects and dynamical effects.

The radiative heating and cooling is the primary driving force of the atmospheric dynamics on tidally locked planets. In principle, the three-dimensional profiles of heating and cooling (e.g., the strength and pressure-dependence of the dayside heating and nightside cooling) are affected by opacities and therefore composition. As noted before, we do not investigate the opacity effect in this study. In order to isolate the effect of composition on the opacity from the other bulk composition effects, the simulations in this study share the same reference properties such as atmospheric opacity, planetary effective temperature, planetary mass, planetary radius, the same distance from the same host star and same cooling mechanism, but only differ in bulk compositions. The temperature change due to radiative heating and cooling also depends on the molecular weight and molar heat capacity (Eq. 1d). A higher molecular weight implies less molecules at a given pressure, resulting a greater temperature change rate, or equivalently a smaller radiative timescale. A higher molar heat capacity suggests that the atmosphere is more difficult to be heated and cooled, leading to a larger radiative timescale. For example, the radiative timescale in a hydrogen atmosphere is more than 15 times longer than the $CO_2$ atmosphere solely due to the effects of $\mu$ and $c_p$.

The dynamical effect is also significant. A typical dynamical timescale in the atmosphere is the wave propagation timescale $\tau_{wave} \sim L/NH$ (Showman et al. 2013), where $L$ is a typical horizontal length scale (e.g., radius of the planet), $N$ is the buoyancy frequency (Brunt-Väisälä frequency) of the atmosphere, and $H = RT/\mu g$ is the pressure scale height. $NH$ approximates the horizontal phase speed of long-vertical-wavelength gravity or Kelvin waves, which is greatly influenced by the molecular weight and molar heat capacity:

$$NH = \left[\frac{RT}{\mu}\left(\frac{R}{c_p} - \frac{\partial \ln T}{\partial \ln p}\right)\right]^{1/2}. \qquad (2)$$

The gravity wave speed $NH$ inversely scales as $\mu^{1/2}$, implying that higher molecular-mass atmospheres have slower gravity/Kelvin wave speeds than lower-molecular-mass atmospheres. For an isothermal atmosphere ($\partial \ln T/\partial \ln p = 0$), $NH$ inversely scales as $c_p^{1/2}$, implying that atmospheres whose molecules have more degrees of freedom likewise have slower gravity wave speeds. Nevertheless, because $c_p$ varies over a smaller range than $\mu$ for plausible compositions (Fig. 1), in practice, the effect of $c_p$ is generally secondary to the effect of $\mu$ on the wave speeds. An exception occurs if the temperature profile is close to an adiabat, because in that situation, $R/c_p$ is close to $\partial \ln T/\partial \ln p$ in Eq. (2). Small variations in $c_p$ can cause large fractional variations in the factor in square brackets, and therefore large fractional changes to the wave speeds and possibly the entire dynamic regime. We will see this effect in Section 5.

In an isothermal atmosphere, the wave propagation timescale $\tau_{wave}$ can be estimated:

$$\tau_{wave} \sim L/NH = \frac{L}{R}\left(\frac{\mu c_p}{T}\right)^{1/2} \qquad (3)$$

showing that the typical dynamical timescale of the atmosphere increases with molecular weight and molar heat capacity.

A useful atmospheric dynamic length scale is the Rossby deformation radius, within which the atmospheric flow is more significantly influenced by gravity and buoyancy effects instead of planetary rotation. In the equatorial region where the Coriolis parameter is small, the Rossby deformation radius ($L_e$) can be approximated as:

$$L_e \sim \left(\frac{NH}{\beta}\right)^{1/2} \sim \left[\frac{RT}{\beta^2\mu}\left(\frac{R}{c_p} - \frac{\partial \ln T}{\partial \ln p}\right)\right]^{1/4}. \qquad (4)$$

The meridional gradient of the Coriolis parameter $\beta = \partial f/\partial y = 2\Omega \cos\phi/R_p$ for planetary radius $R_p$ and latitude $\phi$. For a given temperature-pressure profile, the deformation radius always decreases as $\mu$ and $c_p$ increase. A smaller Rossby deformation radius implies that the typical dynamical features in the system, such as the stable vortex size, are smaller.

The interaction between the radiative effect and dynamical effect is complicated and often leads to noticeable effects on atmospheric dynamics. Showman et al. (2013) showed that hot Jupiters should experience two distinct circulation regimes depending on the relative importance of wave propagation and radiative relaxation. If the radiative timescale is sufficiently short compared with wave propagation timescale, the waves will be damped and jet formation will be inhibited (Showman & Polvani 2011; Showman et al. 2013; Tsai et al. 2014; Zhang & Showman 2014). Therefore the general circulation pattern in the atmosphere is dominated by a substellar-to-antistellar (day-to-night) divergent flow pattern instead of the equatorial zonal jet. In this situation, there is a large day-night temperature difference and small atmospheric hot spot shift in the thermal phase curve. On the other hand, if the radiative timescale is sufficiently long compared with wave propagation timescale, the atmospheric flow will self-organize to form a zonal jet pattern. If we assume the radiative timescale increases with pressure, as generally it does (e.g., Iro et al. 2005; Showman et al. 2008), the general circulation pattern in the atmosphere is expected to exhibit a "regime shift" from a divergent flow in the upper atmosphere to a jet-dominated regime in the deeper atmosphere. Just such a transition in fact occurs in GCM simulations of hot Jupiters (e.g., Cooper & Showman 2005; Showman et al. 2008; Showman et al. 2009; Rauscher & Menou 2010; Heng et al. 2011). An intuitive measure of this transition is the ratio of radiative timescale versus the wave propagation timescale: $\tau_{rad}/\tau_{wave}$, and both the molecular weight and molar heat capacity play important roles in this ratio.

| Experiment | Number of Simulations | Description |
|---|---|---|
| I | 5 | standard cases: $\mu$ and $c_p$ from five typical constituents: $H_2$, He, $H_2O$, $N_2$ and $CO_2$. |
| II | 7 | keep $c_p$ as $7R/2$ but vary $\mu$ of 2, 4, 8, 16, 24, 32 and 44 g $mol^{-1}$, hereafter "mass-2", etc. |
| III | 7 | keep $\tau_{rad}$ same as the "mass-2" case but vary $\mu$ of 2, 4, 8, 16, 24, 32, and 44 g $mol^{-1}$. |
| IV | 4 | keep $\mu$ as 16 g $mol^{-1}$ but vary $c_p$ of $7R/2$, $10R/2$, $12R/2$ and $14R/2$. |

## 3. NUMERICAL AND GENERAL RESULTS

To quantitatively understand the above effects on a tidally locked planet, we simulate the atmospheric circulation using a three-dimensional (3D) general circulation model. Previous studies (e.g., Thrastarson & Cho 2010, Polichtchouk & Cho 2012, Mayne et al. 2014, Cho et al. 2015) demonstrated that the simulation results might be sensitive to the specifications such as forcing and dissipation setup, initial condition and numerical schemes used in different dynamical cores. In this study, we adopted the MITgcm (Adcroft et al. 2004) dynamical core that solves the primitive equations using the finite volume method in a cubesphere grid. MITgcm has been extensively used in studying atmospheric dynamics on planets in and out of the solar system (e.g., Adcroft et al. 2004; Lian & Showman 2010; Showman et al. 2009; Kataria et al. 2014). A detailed investigation by Liu & Showman (2013) on tidally locked planets showed that, under their prescription for forcing and damping, the equilibrated state of the model does not exhibit significant sensitivity to initial conditions. The similar conclusion was also reached by Cho et al. (2015) when both Newtonian cooling and Rayleigh drag are presented in the lower region to dissipate the hot-Jupiter system.

For simplicity, we consider a planet with zero obliquity and in a circular orbit around its host star. We assume the radius, mass and rotation period of the idealized planet are 2.7 Earth radii, 6.5 Earth mass, and $1.365 \times 10^5$ s, respectively. Those properties are similar to GJ1214b. The horizontal resolution of our simulations is C32 in a cubesphere grid, corresponding to $128 \times 64$ in longitude and latitude, respectively. We use 40 levels from about 10 Pa to $10^8$ Pa, evenly spaced in log pressure. We use a time step of 15 second. We apply a fourth-order Shapiro filter to the time derivatives of horizontal velocities and potential temperature with a damping time-scale of 25 second to ensure the numerical stability. The simulations were performed for at least 4000 Earth days in model time so that the statistical properties we will analyze and discuss in this study have reached the equilibrated state. Liu & Showman (2013) has showed that for typical tidally locked exoplanet configurations, the C32 resolution runs exhibit almost the same results as C64 ($256 \times 128$) and C128 ($512 \times 256$). We tested several cases in this study in the C64 grid and 80 levels and the results are consistent with our lower-resolution runs. Finally, all the variables were averaged over the last 350 days in our analysis.

We adopt Newtonian cooling approximation as a simple radiative heating and cooling scheme. The heating term in Eq. (1d) is formulated as:

$$\frac{q\mu}{c_p} = \frac{T_{eq} - T(t)}{\tau_{rad}}. \tag{5}$$

If GJ1214b has a hydrogen-dominated atmosphere, we estimated the radiative timescale $\tau_{rad,H_2}$ on the order of $10^5$ s based on a simple gray cooling scheme:

$$\tau_{rad,H_2} \sim \frac{p_{\tau=1}}{\mu g} \frac{c_p}{4\sigma T^3} \tag{6}$$

where $p_{\tau=1}$ is the pressure level at which the opacity of the atmosphere reaches unity. Motivated by earlier studies using Newtonian cooling (Liu & Showman 2013, Komacek & Showman 2016), we assume that the vertical profile of radiative timescale increases as a power law of pressure, ranging from $10^4$ s at 10 Pa to $10^7$ s at $10^8$ Pa (Fig. 2):

$$\tau_{rad,H_2}(p) = \begin{cases} 10^4 & p < 10^2 \text{ Pa} \\ 10^{5/2} p^{3/4} & 10^2 \text{ Pa} < p < 10^6 \text{ Pa} \\ 10^7 & p > 10^6 \text{ Pa} \end{cases} \tag{7}$$

As noted before, we assume the same opacity for all compositions, implying that $p_{\tau=1}$ is constant. Based on Eq. (6), the radiative timescale $\tau_{rad}$ for other compositions can be scaled based on molecular weight and molar heat capacity:

$$\tau_{rad}(p) = \tau_{rad,H_2}(p) \left(\frac{c_p}{7R/2}\right) \left(\frac{2}{\mu}\right). \tag{8}$$

The spatial distribution of the radiative equilibrium temperature $T_{eq}$ is:

$$T_{eq}(\lambda, \phi, p) = \begin{cases} T_n(p) + \Delta T_{eq}(p) \cos\lambda \cos\phi & \text{dayside} \\ T_n(p) & \text{nightside} \end{cases} \tag{9}$$

where $\lambda$ is longitude and we assume a homogeneous equilibrium temperature in the nightside $T_n(p) = T_0(p) - \Delta T_{eq}(p)/2$. The equilibrium temperature difference $\Delta T_{eq}$ is 600 K at the top of the atmosphere and decreases towards zero at the bottom in a linear fashion with $\ln p$. The mean temperature $T_0$ is calculated using a simplified radiative scheme from Guillot (2010) (Fig. 2). The vertical temperature profile is designed to be stably stratified, meaning that its vertical gradient, $\partial \ln T_{eq}/\partial \ln p$, does not exceed $R/c_p$ for all compositions.

We assume a linear drag scheme in the momentum equation (Eq. 1a), $F_d = \vec{u}/\tau_{drag}$. The drag coefficient ($1/\tau_{drag}$) is assumed as $10^{-7} s^{-1}$ at the bottom of the domain ($10^8$ Pa) and decreases linearly with decreasing pressure to zero at $10^6$ Pa, above which the atmosphere is essentially drag-free. For simplicity, we assume in each simulation that the value of $c_p$ is a constant (using the value at 600 K, Table 1). Thus, we are neglecting the temperature dependence of the heat capacity. In most cases we expect this temperature dependence to be a secondary effect, so this should be a reasonable assumption for the present study. Our tests show that this assumption is actually sufficient to diagnose the role of $c_p$ on the atmospheric dynamics.

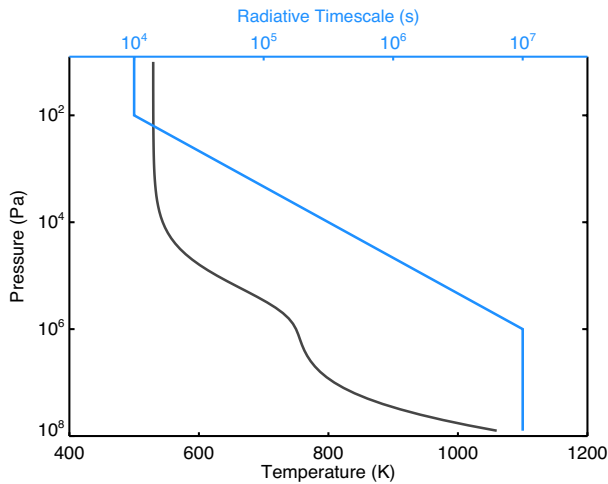

FIG. 2.— Vertical profiles of the mean temperature (black) and a reference radiative timescale (blue) for hydrogen atmosphere ($c_p = 7R/2$ and $\mu = 2$ g mol$^{-1}$).

We specifically designed four sets of experiments to investigate the effects and mechanisms. Table 2 summarizes our simulation experiments. The Experiment I includes the standard cases with five typical constituents: $H_2$, He, $H_2O$, $N_2$ and $CO_2$. In Experiment II, we keep $c_p$ fixed and vary $\mu$ to study the molecular weight effect at constant molar heat capacity. The Experiment III is very similar to II but we fix the radiative timescale $\tau_{rad}$ to separate the radiative effect from dynamical effect of $\mu$. The Experiment IV, in which we keep $\mu$ constant and vary $c_p$ in the simulations, is designed to investigate the molar heat capacity effect at constant molecular weight. The radiative timescale $\tau_{rad}$ in I, II and IV is allowed to vary between simulations but it is fixed in III.

The standard simulations (Experiment I) exhibit important features and trends in the longitude-latitude temperature and wind maps and the zonal-mean zonal wind ($\overline{u}$) distributions (Fig. 3)[1]. The temperature and wind maps from Experiment II (Fig. 4) look very similar to that from Experiment I, implying that the molecular weight effect dominates over the molar heat capacity effect. In order to tie the simulation results to the observations and better understand the theoretical trends, we define several quantities: the day-night temperature contrast, phase offsets of flux peak before the time of secondary eclipse in the thermal phase curve, root-mean-square (RMS) of the wind speed $U_{rms}$, averaged zonal-mean zonal wind speed in the equatorial region $U_z$, and the pressure where the zonal-mean zonal jet speed reaches a maximum (hereafter the "jet core" region). The above quantities exhibit the same trends in Experiments I and II as a function molecular weight from $H_2$ to $CO_2$ (Fig. 5). In the following, we summarize the major trends and qualitatively describe the possible mechanisms.

(1) The temperature contrast at the same pressure level between the dayside and nightside increases with molecular weight (Fig. 4). We adopted the normalized day-night temperature difference ($A$) on a global scale, similar to that in Perez-Becker & Showman (2013) and Komacek & Showman (2016):

$$A(p) = \frac{3}{2\pi} \int_{-\pi/3}^{\pi/3} \left( \frac{\overline{[T(\lambda, \phi, p) - \overline{T}(\phi, p)]^2}}{\overline{[T_{eq}(\lambda, \phi, p) - \overline{T}(\phi, p)]^2}} \right)^{1/2} d\phi. \tag{10}$$

$A \sim 0$ represents homogenous longitudinal temperature distribution and $A \sim 1$ represents a temperature distribution close to the radiative equilibrium temperature pattern. The normalized temperature difference varies from 0.2 in the $H_2$ case to about 1 in the $CO_2$ case (Fig. 5). As the molecular weight increases, the radiative timescale decreases and the dynamical timescale (wave propagation timescale) increases, leading to a more efficient relaxation of the thermal field towards the radiative equilibrium state and increase of the day-night temperature contrast. A smaller wind speed in higher-molecular-weight atmosphere (Fig. 5) also acts less efficiently to redistribute heat and smooth the thermal field by advection.

(2) The phase offset before the secondary eclipse, which corresponds to a hot spot in the atmosphere shifted east of the sub-stellar point (Fig. 4, left column), decreases with increasing molecular weight (Fig. 5). As discussed in trend (1), the decrease of the radiative timescale and the zonal wind speed with molecular weight helps maintain the radiative equilibrium thermal field, weakening the ability of advection to distort the temperature field.

(3) The RMS wind speed $U_{rms}$ decreases with molecular weight (Fig. 5). Here we calculated $U_{rms}$ based on horizontal wind speeds $u$ and $v$ weighted by area over a 120 degree latitude band at each pressure level:

$$U_{rms}(p) = \left( \frac{1}{\sqrt{3}} \int_{-\pi/3}^{\pi/3} \overline{u(\lambda, \phi, p)^2 + v(\lambda, \phi, p)^2} \cos \phi d\phi \right)^{1/2}. \tag{11}$$

However, this decreasing trend is less related to the changes of radiative and dynamical timescales. A smaller $U_{rms}$ in the $CO_2$ atmosphere than the $H_2$ atmosphere is more associated with that the atmospheric scale height is smaller in the $CO_2$ atmosphere than that in the $H_2$ atmosphere. An atmosphere composed of heavier molecules will have a smaller day-night pressure-gradient force, leading to a smaller wind field in general. A more rigorous discussion on the wind speed will be presented in Section 4.3.

(4) The latitudinal width of the equatorial jet decreases with increasing molecular weight (Fig. 3). Fig. 5 presents the e-folding width of the equatorial zonal jet as function of molecular weight. In the $H_2$ case, the zonal-mean zonal wind is dominated by a broad eastward equatorial jet, which, over at least certain ranges of pressure, extends nearly to the poles. The jet width shrinks with increasing molecular weight. Fig. 3 shows that, for the case of $H_2O$, the equatorial jet is sufficiently narrow to allow room for broad regions of westward flow at mid-to-high latitudes. For $N_2$ and $CO_2$, the equatorial jet is even narrower, and and two new high-latitude eastward wind regions appear. The same trends emerge in Fig. 4 when molecular mass is varied while holding molar heat capacity constant. This trend is related to the decrease of dynamical length scale with increasing molecular weight.

---

[1] As a notation, we use the overbar to denote a zonal average. A quantity $A$ can be decomposed as $A = \overline{A} + A'$, where $\overline{A}$ is the zonal-mean value averaged over a latitude circle and $A'$ is the local departure from the zonal average or so-called "eddy" value.

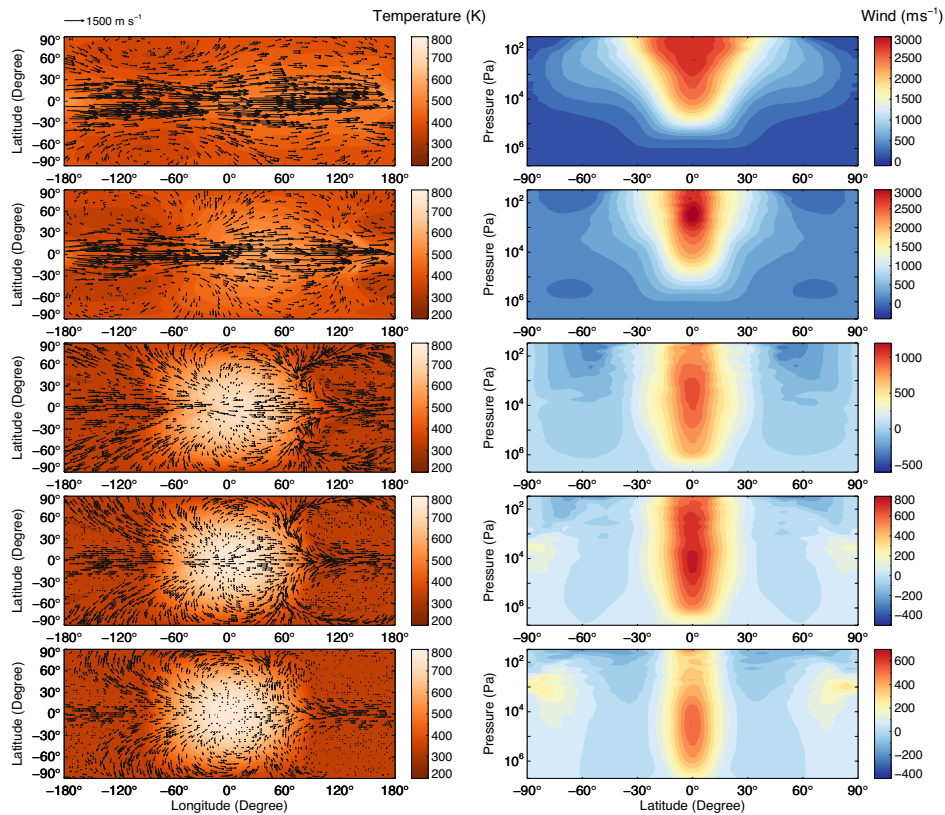

FIG. 3.— Longitude-latitude temperature and wind maps at 700 Pa (left column) and the zonal-mean zonal wind ($\overline{u}$) distributions (right column) as a function of pressure and latitude for the typical constituents (Experiment I). From top to bottom: H$_2$, He, H$_2$O, N$_2$ and CO$_2$.

(5) At the same pressure, the general circulation pattern gradually shifts from the zonal-jet-dominated regime to the day-to-night circulation regime as the molecular weight increases (Fig. 4, left column). Correspondingly, the jet core pressure occurs deeper in the atmosphere (right column of Fig. 4, Fig. 5). This is closely related to the interplay between the radiative timescale and wave propagation timescale. Showman et al. (2013) showed that when the radiative timescale is significantly less than the wave-propagation timescale, the atmosphere tends to be dominated by a day-night flow. Whereas strong zonal jets emerge when the radiative timescale is comparable to or longer than the wave propagation timescale. Because the ratio of the radiative timescale to the wave propagation timescale $\tau_{rad}/\tau_{wave}$ decreases with increasing molecular weight, the transition between the day-to-night circulation regime and the jet-dominated regime moves deeper into the atmosphere.

(6) We defined the root-mean-square of the equatorial zonal-mean zonal wind speed $U_z$ and the eddy compo-

nent $U_e$ over a 60 degree latitude band:

$$U_z(p) = \left( \int_{-\pi/6}^{\pi/6} \overline{u(\phi, \lambda, p)}^2 \cos\phi \, d\phi \right)^{1/2} \quad (12)$$

$$U_e(p) = \left( \int_{-\pi/6}^{\pi/6} \overline{u'(\phi, \lambda, p)^2} \cos\phi \, d\phi \right)^{1/2} \quad (13)$$

If the zonal velocity $u$ is much larger than the meridional velocity $v$, $U_{rms}$ in the equatorial domain (Eq. 11, but averaged in the equatorial region) can be related to $U_z$ and $U_e$ via $U_{rms}^2 \approx U_z^2 + U_e^2$. The maximum zonal-mean zonal wind $U_z$ in the atmosphere ("jet core speed") decreases with molecular weight. This might be associated with the trends (3) and (5).

The molar heat capacity has a secondary effect, which can be seen from the small deviation between the typical composition simulations and the cases with molar heat capacity fixed (Fig. 5), for instance, the small departure between the He case and the case with the same molecular weight (4 g mol$^{-1}$) but $c_p$ fixed (hereafter "mass-4"), and also the difference between the CO$_2$ case and the mass-44 case. More discussion will be presented in Section 5.

Next we will present analytical theories and quantitatively discuss each of those trends. In order to elaborate the dependence of these trends on the molecular weight more clearly, our discussion will mainly focus on the Experiment II.

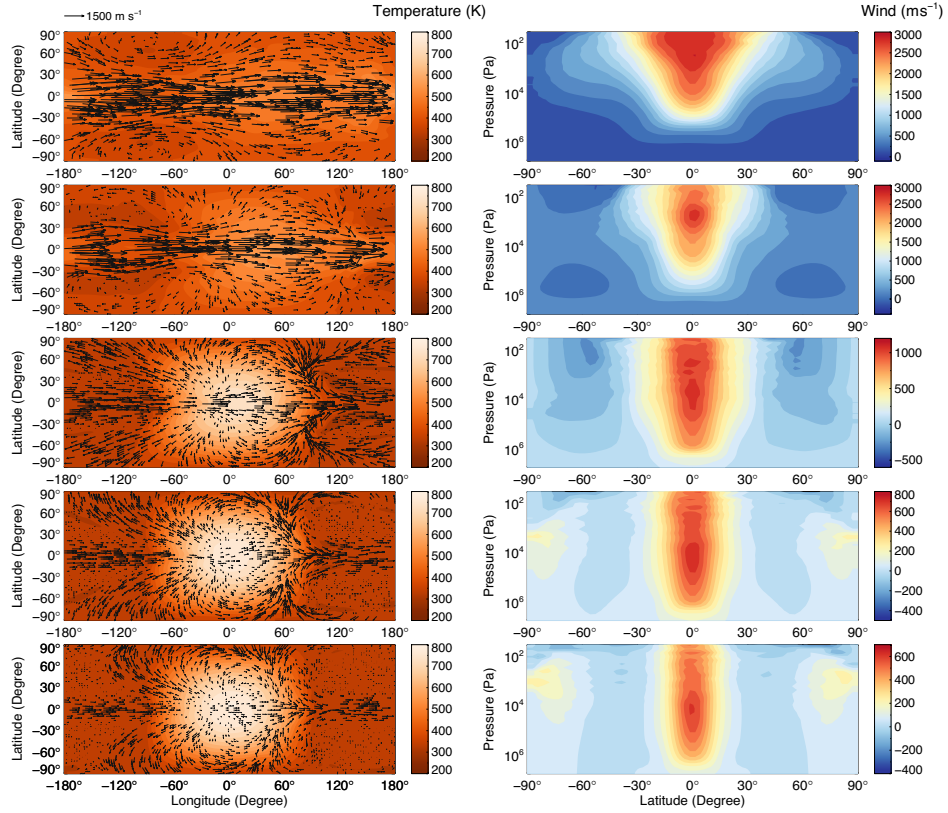

Fig. 4.— Longitude-latitude maps of temperature and wind at 700 Pa (left column) and the zonal-mean zonal wind ($\bar{u}$) distributions (right column) as a function of pressure and latitude from Experiment II. From top to bottom: the molecular weight of 2, 4, 16, 32 and 44 g mol$^{-1}$, respectively.

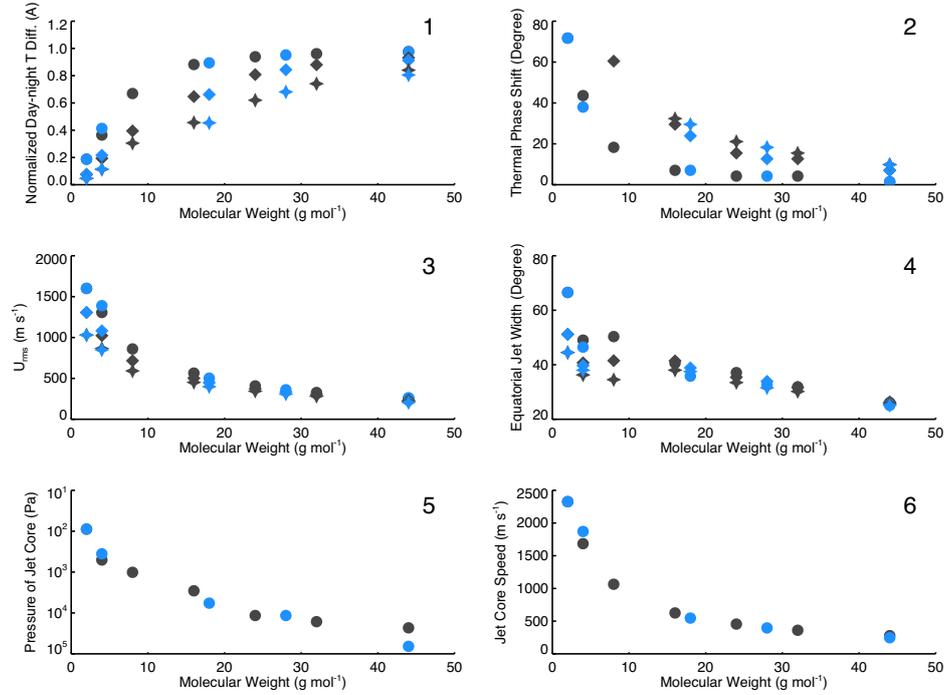

Fig. 5.— Summary of the major trends 1 to 6 as a function of molecular weight for Experiment I (blue) and II (black). The trends 1 to 4 are computed at three pressure levels: 700 Pa (dots), 5700 Pa (diamonds) and 23100 Pa (stars).

## 4. EFFECTS OF MOLECULAR WEIGHT

### 4.1. *Trend 1: Day-night Temperature Contrast*

The dayside-nightside temperature difference is a fundamental aspect of the overall circulation and moreover directly controls the amplitude of the phase variation in infrared phase curves—a major observable. Our simulations show that the day-night temperature difference increases with molecular weight (Fig. 3-4). This results from the effect of the molecular weight on both the dynamics and on the radiative timescale via Eq. (5). To separate these effects, Experiment III shows how the day-night temperature difference varies with molecular mass when the radiative timescale is held fixed as the hydrogen case (Eq. 7). The longitude-latitude maps of temperature are shown in Fig. 6. The integrated dayside-nightside temperature difference $A$, shown in Fig. 7, indicate that even when radiative time constant is held fixed, the day-night temperature difference increases with molecular mass—though more weakly than in the Experiment II.

Here, we discuss the physical reasons for all of these trends and compare them to theory. In particular, Perez-Becker & Showman (2013) and Komacek & Showman (2016) presented analytical theories for the day-night temperature difference on hot Jupiters; Perez-Becker & Showman (2013) constructed their theory for a 1.5-layer shallow-water model, and Komacek & Showman (2016) extended it to the three-dimensional primitive equations. Here, we show that the Komacek & Showman (2016) theory explains our simulated trends well. We also present a modified form of the Komacek & Showman (2016) theory that, while incorporating the same assumptions, is more analytically compact.

Perez-Becker & Showman (2013) and Komacek & Showman (2016) pointed out that the day-night temperature difference is controlled by several intrinsic timescales in the atmosphere, including the rotational timescale, radiative timescale, wave propagation timescale, horizontal and vertical advection timescale, and the frictional drag timescale (if drag is present). By combining the horizontal momentum equation and the thermodynamic energy equation under the "Weak Temperature Gradient" (WTG) approximation (Sobel et al. 2001), we can obtain analytical expressions for a proxy of day-night temperature difference ($A = \Delta T / \Delta T_{eq}$) and the RMS wind velocity $U_{rms}$, on synchronously rotating planets. In the momentum equation, the day-night pressure gradient force, which results from the day-night temperature difference and is responsible for driving the circulation, is primarily balanced against the largest of the Coriolis, horizontal-advection, vertical-advection, and frictional-drag forces per mass, respectively. The four possible balances lead to four distinct expressions for $A$, along with conditions that determine which of these four regimes applies in any given portion of the parameter space (Komacek & Showman 2016). This separation into regimes is helpful for clarifying our understanding of the behavior in each regime. Nevertheless, it would be useful to obtain a single analytic expression for $A$ that interpolates smoothly between the regimes. Here we united the above four regimes and achieved a single and close-form expression of $A$ in Appendix A:

$$A = \frac{\Delta T}{\Delta T_{eq}} \sim 1 - \frac{2}{\alpha + \sqrt{\alpha^2 + 4\gamma^2}} \qquad (14)$$

where the non-dimensional parameters $\alpha$ and $\gamma$ are defined as:

$$\alpha = 1 + \frac{(\Omega + \frac{1}{\tau_{drag}})\tau_{wave}^2}{\tau_{rad}\Delta\ln p} \qquad (15)$$

$$\gamma = \frac{\tau_{wave}^2}{\tau_{rad}\tau_{adv,eq}\Delta\ln p} \qquad (16)$$

Here, $\tau_{drag}$ is the frictional drag timescale. $\tau_{wave} = L/NH$ is the timescale for horizontal wave propagation. The quantity $\tau_{adv,eq} = L/U_{eq}$ is a reference advective timescale due to the "equilibrium cyclostrophic wind" $U_{eq} = (R\Delta T_{eq}\Delta\ln p/2\mu)^{1/2}$, which is a hypothetical, reference wind speed that would result from acceleration of the wind from day to night due to the day-night pressure gradient if the day-night temperature difference were in radiative equilibrium and if the Rossby number exceeds unity (compare to Showman et al. 2010, Eq. 48). In the above, $\Delta\ln p$ is the difference in log pressure between some deep pressure where the day-night temperature difference is small (10 bars in the theory and simulations from Komacek & Showman 2016) and some smaller pressure of interest in the observable atmosphere. See Komacek & Showman (2016) for more details about the motivations for and assumptions underlying the theory, and comparisons with a grid of GCM simulations for hot Jupiters covering a wide range of $\tau_{rad}$ and $\tau_{drag}$.

This theory implies that the day-night temperature contrast is smaller if the radiative timescale is larger or the dynamical timescale is smaller. To view the dependence on molecular weight, we evaluate each term in $\alpha$ and $\gamma$. Assuming the dynamical length scale $L$ is constant (representing a planetary radius associated with the day-night temperature contrast, for example), then $\tau_{wave} \propto \mu^{1/2}$ in an isothermal atmosphere (Eq. 3) and the advective timescale $\tau_{adv,eq}$ is scaled as $\mu^{1/2}$. If $\tau_{rad}$ is considered constant, then the second term in the expression for $\alpha$ (Eq. 15) is proportional to $\mu$, and $\gamma \propto \mu^{1/2}$ (Eq. 16). Whereas if $\tau_{rad}$ scales inversely with $\mu$ as expected from Eq. (8), then the second term in the expression for $\alpha$ is proportional to $\mu^2$ and $\gamma \propto \mu^{3/2}$. Therefore, Eq. (14) predicts that the day-night temperature contrast should increase with the molecular weight, and this dependence should be weaker in the case where $\tau_{rad}$ is constant than in the case where $\tau_{rad}$ decreases with increasing molecular weight. Both predictions are qualitatively consistent with the trends in our simulations (Fig. 7).

This theory implies that in the limit of very high molecular mass ($\mu \to \infty$), the atmospheric temperature structure tends toward radiative equilibrium. This can be seen from the fact that, in this limit, the denominator in the second term on the righthand side of Eq. (14) becomes infinite, which therefore implies that $A \to 1$. The convergence toward radiative equilibrium as $\mu \to \infty$ is fastest if we allow $\tau_{rad}$ to scale inversely with $\mu$ (Eq. 8) but occurs even if $\tau_{rad}$ were taken to be independent of molecular weight.

What is the physical reason for the convergence toward radiative equilibrium at high molecular mass? Perez-Becker & Showman (2013) and Komacek & Showman (2016) showed that the day-night temperature difference is regulated by a wave-adjustment mechanism: planetary-scale waves that propagate from the dayside to the nightside act to adjust isentropes up or down in an attempt to flatten them, which, acting in isolation, attempts to produce a state with minimal day-night temperature differences. This process operates ef-ficiently when the waves can propagate from the dayside to the nightside faster than radiation or friction can act to damp them. On the other hand, if friction or radiation damps the waves before they propagate from the dayside to the nightside, then the wave-adjustment mechanism is suppressed, and the day-night temperature difference becomes large. In the limit of high molecular mass ($\mu \to \infty$), the wave speeds $NH$ become slow, and therefore the day-night wave propagation timescale $\tau_{wave}$ becomes long (Eq. 3)—which gives the atmosphere ample time to respond to radiation for any plausible, finite value of $\tau_{rad}$. In other words, the limit $\mu \to \infty$ corresponds to the limit where the wave-adjustment mechanism acts less and less efficiently, allowing it to be easily suppressed.

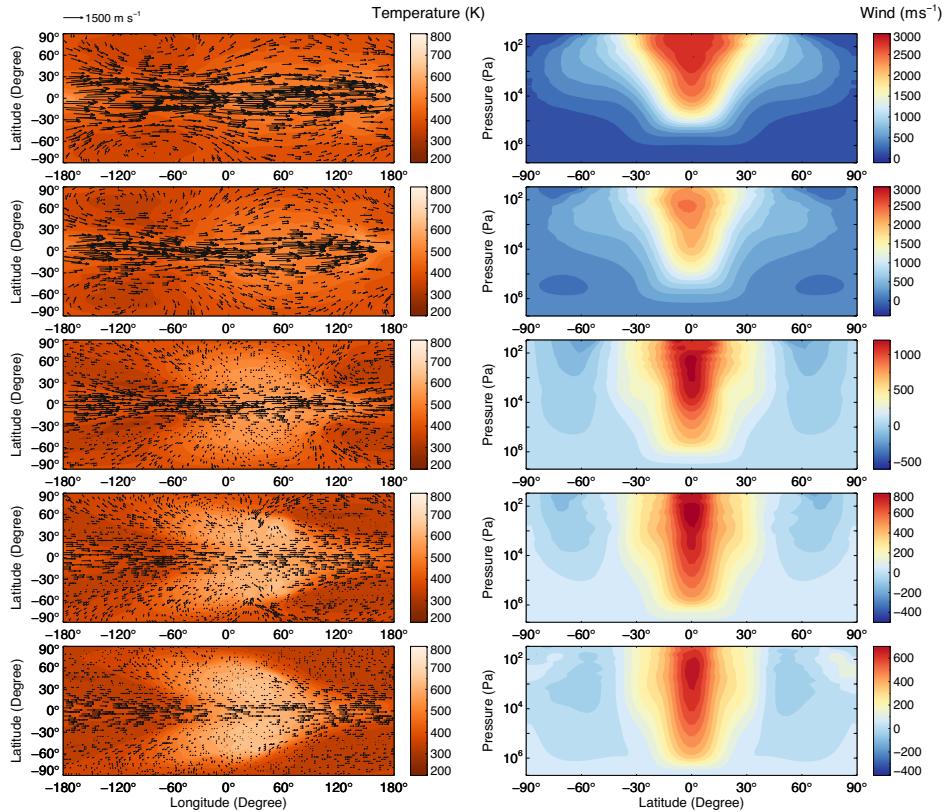

FIG. 6.— Longitude-latitude maps of temperature and wind at 700 Pa (left column) and the zonal-mean zonal wind ($\bar{u}$) distributions (right column) as a function of pressure and latitude from Experiment III in which the radiative timescale $\tau_{rad}$ is fixed. From top to bottom: the molecular weight of 2, 4, 16, 32 and 44 g mol$^{-1}$, respectively.

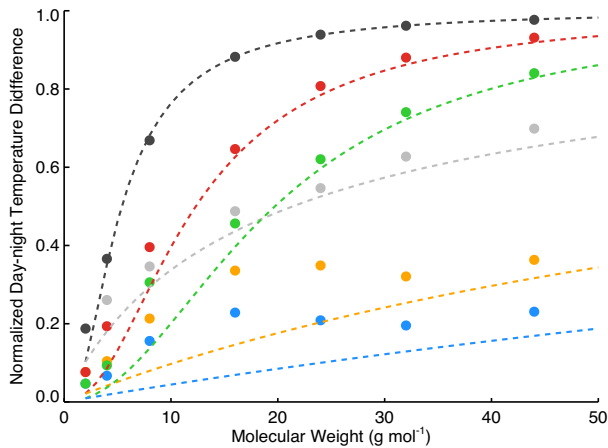

FIG. 7.— Normalized temperature difference between the dayside and nightside as a function of molecular weight from Experiment II (black for 700 Pa, red for 5700 Pa and green for 23100 Pa) and Experiment III (gray for 700 Pa, orange for 5700 Pa and blue for 23100 Pa). Dashed lines are the prediction from Eq. (14).

Eq. (14) also quantitatively predicts the day-night temperature difference from first principles. In our study, drag is not important above the bottom frictional layer (inviscid limit) therefore $\tau_{drag}$ can be neglected. We use the mean temperature profile $T_0$ (Fig. 2) to estimate $\tau_{wave}$. At 700 Pa, taking the horizontal length scale as $L \sim R_p = 1.71 \times 10^7$ m, $T_0 \sim 550$ K, $\Delta T_{eq} \sim 500$ K, and radiative timescale based on Eq. (8), we obtain the theoretical curve of $A$ as function of molecular weight (Fig. 7). The predictions agree fairly well with our simulation results. A comparison of the theory with the simulations at other pressure levels demonstrates good agreement as well (Fig. 7).

The theory also predicts that $A$ should decrease with increasing pressure. Among the parameters in $\alpha$ and $\gamma$,

$\tau_{rad}$ increases by several orders of magnitude from the top to the bottom of the atmosphere, while the other parameters such as $\tau_{wave}$, $\tau_{adv,eq}$ and $\Delta \ln p$ have much weaker pressure dependence. From Eq. (14), $A$ should monotonically decrease with pressure. Since $\Delta T_{eq}$ also decreases with increasing pressure, the day-night temperature contrast should be smaller in the deeper atmosphere, consistent with our simulation results.

We also investigated the significance of the radiation effect versus the dynamical effect. In Experiment III, we fixed the radiative timescale for all simulations regardless of molecular weight. The resulting day-night temperature difference is much lower than the simulations where $\tau_{rad}$ changes with $\mu$ (Fig. 7), suggesting an important role of the radiative relaxation. For the analytical prediction, we use a constant radiative timescale $\tau_{rad} \sim 4.38 \times 10^4$ s but keep the same previous parameters, the theoretical prediction agrees reasonably well with the Experiment III simulations (Fig. 7).

### 4.2. Trend 2: Thermal Phase Curve and Phase Shift

Due to the strong equatorial prograde jet, the temperature maximum is advected eastward with respect to the sub-stellar point, causing the peak of thermal flux to become shifted before the secondary eclipse. This phase offset was first predicted by Showman & Guillot (2002) and later confirmed by thermal phase curve observations (e.g., Knutson et al. 2007). Cowan & Agol (2011) systematically investigated the relationship between the horizontal advection and the thermal phase curve in a semi-analytic model. With a 3D GCM, Menou (2011) also found a significant decrease of the ratio between advection timescale and radiative timescale as the metallicity decreases, leading to a large difference in the thermal phase curves.

Based on our simulations, for each pressure level, we first integrated the emergent thermal flux over the face-on hemisphere as a function of phase ($\delta$) as the planet orbits/rotates:

$$F(\delta, p) = \int_{\pi/2-\delta}^{3\pi/2-\delta} d\lambda \int_{-\pi/2}^{\pi/2} \sigma T^4(\lambda, \phi, p) R_p^2 \cos^2 \phi \cos \lambda d\phi.$$

(17)

Here, we define $\delta$ such that $\delta = 0°$ corresponds to the time of primary transit (when the planet's nightside faces Earth) and $\delta = 180°$ corresponds to the time of secondary eclipse (when the planet's dayside faces Earth), and $\delta = 90°$ and $\delta = 270°$ correspond, respectively, to the times when the planet's western and eastern terminators lie at the sub-Earth longitude. Note that longitude is defined such that $\lambda = 0°$ is the substellar point and $\lambda = 180°$ is the anti-stellar point.

Then we removed the mean of the integrated flux and obtain a normalized "thermal phase curve" $C(\delta, p)$ at each pressure level ($p$) as a function of orbital phase (in degrees):

$$C(\delta, p) = \frac{F(\delta, p) - \langle F(\delta, p) \rangle}{\langle F(\delta, p) \rangle}$$

(18)

where $\langle F(\delta, p) \rangle$ denotes the averaged flux over the full phase. Note that the "thermal phase curve" defined in this study is slightly different from the realistic observed light curves because we did not perform radiative transfer calculation that includes all pressure levels. Instead, our thermal phase curve defined at pressure level $p$ should be understood as an approximation in which we assume that the weighting function of the outgoing thermal flux at certain wavelength has a sharp peak at pressure level $p$ and therefore the contribution from other pressure levels is neglected.

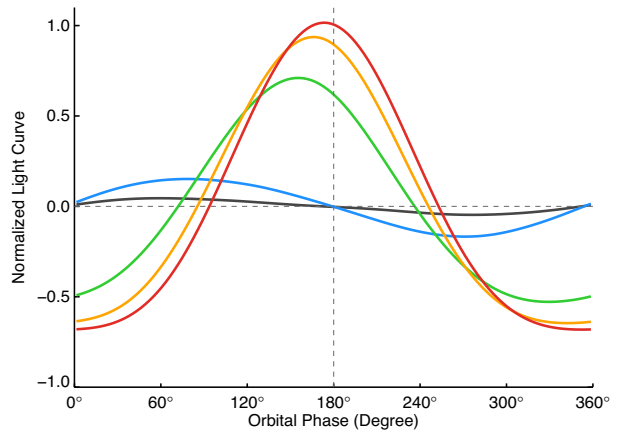

Fig. 8.— Normalized thermal phase curves for Experiment I at 5700 Pa: $H_2$ (black), He (blue), $H_2O$ (green), $N_2$ (orange) and $CO_2$ (red). The primary transit occurs at phase 0° and the secondary eclipse occurs at phase 180°.

The normalized thermal phase curves at 5700 Pa for the standard cases are illustrated in Fig. 8. For clarity, the primary and secondary eclipses are not shown here. There is a clear anti-correlation between the phase offset before the secondary eclipse ($\delta = 180°$) and the amplitude of the phase curve. From hydrogen to $CO_2$, the phase shift is less as the amplitude increases. The thermal phase curve of the $CO_2$ case is almost symmetric about the secondary eclipse, and the amplitude of the maximum flux is actually larger than twice of the mean emergent flux. Although the phase curve is an integrated effect of the temperature distribution, we found that the degree of thermal phase shift ($\delta_s$) with respect to the secondary eclipses can be approximated by longitudinal offset of temperature maximum ($\lambda_m$) with respect to the substellar point in our simulations.

Our previous theory for day-night temperature difference (Eq. 14) only predicts a global-average quantity $A$ but does not provide the longitudinal variation of the temperature, especially the offset of the temperature field. In order to quantitatively investigate the thermal phase shift, we construct in Appendix B a simple kinematic model to study how the longitudinal temperature distribution is influenced by a constant zonal wind. Our procedure is similar to that in Cowan & Agol (2011) but different in the radiative relaxation schemes. To ensure consistency with our 3D simulations, we adopt the Newtonian cooling scheme where the relaxation rate is proportional to temperature, while in Cowan & Agol (2011) the relaxation rate is proportional to $T^3$. In Appendix B we show that the temperature distribution and therefore the thermal phase curve can be characterized by three parameters in the atmosphere: $T_n$, $\Delta T_{eq}$, and $\xi$, where $\xi = \tau_{rad}/\tau_{adv}$ is the ratio between the radiative timescale $\tau_{rad}$ and the advection timescale $\tau_{adv}$. The advection timescale here should be defined as $\tau_{adv} = L/\overline{u}$ for a specified zonal-mean zonal wind $\overline{u}$. Therefore, this is not a first-principle theory because we do not have a theory to predict the zonal-mean zonal wind. With the $\overline{u}$ from numerical simulations, the analytical temperature distributions and thermal phase curves agree fairly well with the 3D GCM results (Appendix B). Our analytical theory provides a simple but physically based tool for future analysis of transit thermal phase curves.

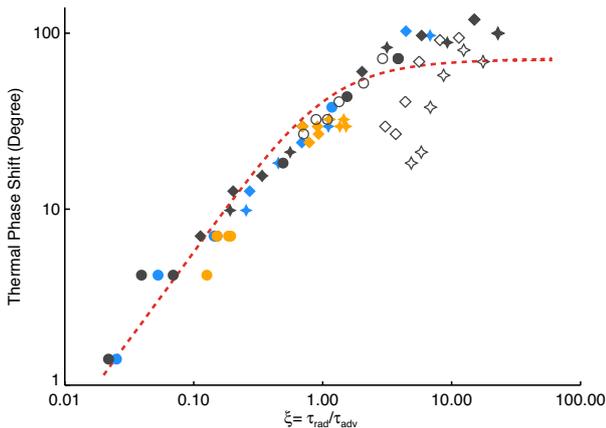

Fig. 9.— Thermal phase shifts (symbols) from all simulation cases and analytical predictions (dashed line) from Eq. (19) as a function of the ratio between the radiative timescale and the advection timescale. Experiments I to IV in Table 2 corresponds to symbols of blue, black filled, black open, orange filled, respectively. Results from three pressure levels are shown here: 700 Pa (dots), 5700 Pa (diamonds) and 23100 Pa (stars).

In Appendix B we also show that the longitudinal hot spot offset $\lambda_m$ satisfies:

$$\sin(\lambda_s - \lambda_m)e^{\lambda_m/\xi} = \frac{\eta}{\xi \cos \lambda_s}. \quad (19)$$

where

$$\lambda_s = \tan^{-1}\xi \quad (20)$$

and

$$\eta = \frac{\xi}{1+\xi^2}\frac{e^{\frac{\pi}{2\xi}} + e^{\frac{3\pi}{2\xi}}}{e^{\frac{2\pi}{\xi}} - 1}. \quad (21)$$

$\lambda_m$ cannot be expressed explicitly but can be solved easily. When $\lambda_m$ is small, $\lambda_m \approx \lambda_s = \tan^{-1}\xi$. Given our expectation that the longitudinal offset of the hot spot ($\lambda_m$, as measured in degrees) is similar to the phase shift of the flux peak in the lightcurve ($\delta_s$, again expressed in degrees), Eq. (19) can thus be used to estimate the phase shift in the lightcurve for a particular model. Note that $\lambda_m$ only depends on $\xi$. It implies the thermal phase shift $\delta_s$ might be mainly dominated by $\xi$ as well (see Appendix B for more discussion). If we take $\tau_{adv} = R_p/\overline{u}$ where $\overline{u}$ is calculated from the simulations (Eq. 12, but averaged from latitude $-60°$ to $60°$), the predicted phase shifts from Eq. (19) agree fairly well with the thermal phase shift in all simulation cases for several pressure levels (Fig. 9). It suggests that our kinematic thermal model has captured the essential physics to explain the zonal distribution of thermodynamic quantities in 3D simulations.

Our simple theory reveals an essential relationship between the phase shift and two fundamental timescales in the system. If the radiative timescale is very short relative to the advective timescale, i.e., $\tau_{rad}/\tau_{adv} \sim 0$, then $\delta_s \sim 0°$, meaning that there is no thermal phase shift.

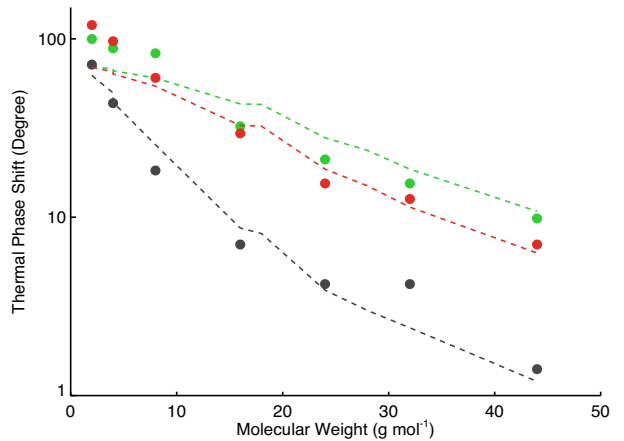

Fig. 10.— Thermal phase curve shift as a function of molecular weight at at 700 Pa (black), 5700 Pa (red) and 23100 Pa (green) from Experiments II. The dashed line is the analytical expression based on Eq. (19).

The atmosphere thermal structure is entirely controlled by the radiation and retains the equilibrium temperature distribution, $T \sim T_{eq}$. On the other hand, if the radiative timescale is very long compared with the advective timescale, then the phase offset shifts toward 90 degrees, and the predicted amplitude of the phase curve variation goes to zero. The increasing trend of thermal phase shift with $\tau_{rad}/\tau_{adv}$ qualitatively agrees with the results in Cowan & Agol (2011) and Menou (2011).

The trend of the horizontal temperature distribution is mainly due to the molecular weight effect, as evidenced by the similarities between Experiment I (Fig. 3) and Experiment II (Fig. 4). The radiative timescale is inversely proportional to the molecular weight (Eq. 9). The advective timescale $\tau_{adv}$ due to the zonal-mean zonal wind increases with molecular weight (Fig. 5). Therefore, the ratio $\tau_{rad}/\tau_{adv}$ must decrease with increasing molecular weight, and correspondingly, the thermal phase shift decreases, as shown in Experiments I and II (Fig. 10). Since $\tau_{rad}$ increases dramatically (Fig. 2) with pressure and $\tau_{adv}$ has a weaker dependence on pressure (suggested by $\overline{u}$ in Fig. 4), the theory implies that the phase shift is larger at greater depth in the atmosphere. Our simulations confirm that the phase shifts increase with pressure (Fig. 9). However, this prediction cannot be applied to the jet core region where the atmosphere is overwhelmingly dominated by jets and waves and the thermal phase curve does not look like a sinusoidal-type curve but a more irregular pattern.

Despite the success of Eq. (19) in explaining the simulation results, we should emphasize that it is not a fully predictive theory, since it can only be evaluated with knowledge of zonal-mean zonal wind speeds. A fully predictive, first principles theory of the hot spot offset would require a quantitatively accurate theory for the zonal-mean zonal wind speed, which is currently lacking.

### 4.3. *Trend 3: Global Wind Speed $U_{rms}$*

The same theory for the day-night temperature difference in Appendix A also predicts the RMS wind speed $U_{rms}$ (hereafter $U$):

$$\frac{U}{U_{eq}} \sim \sqrt{(\alpha/2\gamma)^2 + 1} - \alpha/2\gamma \quad (22)$$

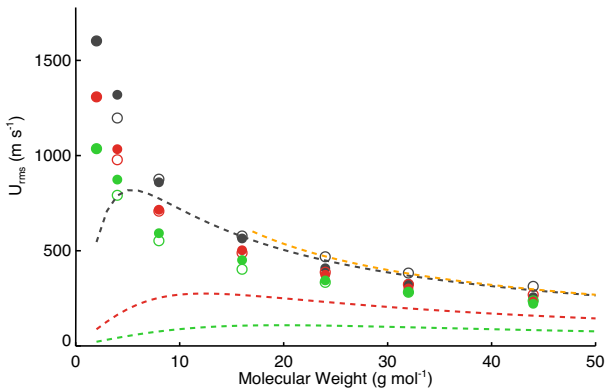

Fig. 11.— RMS wind speeds $U_{rms}$ as a function of molecular weight at 700 Pa (black), 5700 Pa (red) and 23100 Pa (green) from Experiments II (filled circles) and III (open circles). Dashed lines with corresponding colors are the predictions from Eq. (22). The yellow dashed line is the prediction from Eq. (24) at 700 Pa in the limit of large day-night temperature contrast ($A \sim 1$).

where $\alpha$ and $\gamma$ are from Eq. (15) and (16), respectively. Again, this is essentially identical to the prediction from Komacek & Showman (2016) except that it combines their distinct expressions from the multiple regimes (for possible balances in the momentum equation) into a single expression that smoothly varies between the regimes and is valid across the entire parameter space. Recall that the hypothetical "equilibrium cyclostrophic wind" $U_{eq} = (R\Delta T_{eq}\Delta \ln p/2\mu)^{1/2}$ is inversely proportional to the square root of the molecular weight (Section 4.1). The ratio of the wind speed to $U_{eq}$ solely depends on a single parameter $\alpha/2\gamma$.

Our simulations show that the RMS wind speed decreases with increasing molecular weight (Fig. 11). With the same parameters in Section 4.1 for 700 Pa, the predicted wind speeds are in a good agreement with the simulation results in high molecular weight cases but underestimate the low-molecular-weight simulation results (e.g., $\mu < 10$ g mol$^{-1}$) by a factor of 2 to 3 (Fig. 11). The analytical expression of wind speed at a given pressure level generally shows a non-monotonic behavior as a function of molecular weight in the low molecular weight regime, which is not consistent with the numerical simulations. At larger pressure levels, the predicted wind speeds are smaller than the simulations for all molecular weight cases. The agreement is worse in the lower molecular regime. It seems the predicted wind speeds decrease much faster towards deeper atmosphere than that in the numerical simulations. This factor of 2 to 3 discrepancy between model and theory have been seen in hydrogen atmosphere simulations (hot Jupiters) in Komacek & Showman (2016) with large drag timescale and radiation timescale, and Fig. 6 of Perez-Becker & Showman (2013), suggesting some caveats in our analytical scaling theory. The cause has yet to be investigated.

It is worth considering more theoretical details in the limit of high molecular weight. We showed in Section 4.1 that, at high molecular weight limit ($\mu \to \infty$), the temperature converges toward the radiative-equilibrium temperature structure ($A \sim 1$ in Fig. 7). This results from the fact that the wave speeds in the high molecular weight limit become slow, which means that the wave-adjustment mechanism that regulates the day-night temperature differences (Perez-Becker & Showman 2013;

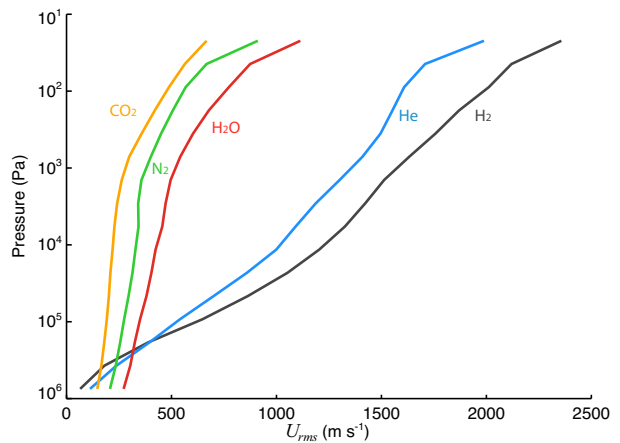

Fig. 12.— The simulated RMS wind speeds $U_{rms}$ decrease with pressure from Experiment I.

Komacek & Showman 2016) becomes easily suppressed. Thus, we can simply replace $\Delta T$ with $\Delta T_{eq}$ in the momentum equation (A.1a in Appendix A). For a drag-free atmosphere, the momentum equation then becomes:

$$\frac{U^2}{L} + \Omega U \sim \frac{R\Delta T_{eq}\Delta \ln p}{2\mu L} \qquad (23)$$

This directly yields wind velocity:

$$\frac{U}{U_{eq}} \sim \sqrt{Ro_{eq}^{-2} + 1} - Ro_{eq}^{-1} \qquad (24)$$

where $Ro_{eq} \equiv 2U_{eq}/\Omega L$ is the Rossby number associated with the "equilibrium cyclostrophic wind" $U_{eq}$. The wind speed predicted by Eq. (24) agrees quantitatively well with the simulations for the high mass cases at 700 Pa (Fig. 11).

Strictly speaking, Eq. (23) and (24) are only valid when $\Delta T = \Delta T_{eq}$, which occurs in the limit where the molecular weight is sufficiently large, i.e., $\mu \to \infty$, so that $U_{eq} \to 0$ and $Ro_{eq} \to 0$, leading to $U = U_{eq}^2/\Omega L \propto \mu^{-1}$ in Eq. (24). In other words, in this limit, since the pressure-gradient force (right hand side of Eq. 23) is sufficiently weak, we expect the winds to be weak—the term quadratic in speed, $U^2/L$, will become smaller than the linear term $\Omega U$. Therefore the inverse dependence of pressure-gradient force on $\mu$ translates directly into an inverse dependence of $U$ on $\mu$.

However, in reality, the horizontal temperature difference $\Delta T$ in the cases with $\mu > 10$ g mol$^{-1}$ at low pressures in Experiment II is already fairly close to the radiative-equilibrium temperature contrast $\Delta T_{eq}$ (Fig. 7). Therefore, Eq. (23) and (24) are also a good approximation for atmospheres with moderate molecular weight at those pressure levels. Based on Eq. (24), if the Rossby number $Ro_{eq} \gg 1$, $U = U_{eq} \propto \mu^{-1/2}$; if $Ro_{eq} \ll 1$, then $U = U_{eq}^2/\Omega L \propto \mu^{-1}$. As the molecular weight increases, the pressure-gradient force deceases, leading to a smaller $U_{eq}$ and smaller Rossby number $Ro_{eq}$. As a result, the dependence of RMS wind speed on molecular weight will gradually shift from the $\mu^{-1/2}$ regime to the $\mu^{-1}$ regime. In our simulations, the wind speed $U$ decreases with $\mu$ in a trend between $\mu^{-1/2}$ and $\mu^{-1}$ (Fig. 11).

The physical reason that the wind speed decreases with increasing molecular mass in the high mass limit is straightforward. In a hydrostatically balanced atmosphere, the pressure-gradient force is essentially the result of a vertical integral over the day-night temperature difference. At high molecular mass, the scale height is small, and thus there is not much altitude difference between the photosphere pressure and the pressure of a given deep isobar where the day-night temperature difference is small. The larger the molecular mass, the smaller is the altitude range over which the day-night temperature difference acts. This leads to a smaller pressure-gradient force, which manifests as the inverse $\mu$ dependence in the right hand side of Eq. (23). As a result, the RMS wind speed, which is the solution of the Eq. (23) and directly driven by the pressure-gradient force, should decrease with increasing molecular weight accordingly. Vertically, because the pressure-gradient force decreases towards deeper into the atmosphere, $U_{rms}$ decreases towards higher pressure for all cases (Fig. 12).

Eq. (24) also shows that, in a drag-free atmosphere, the theoretical wind speeds in the high molecular weight regime are solely controlled by three parameters: the planetary rotation rate, equilibrium day-night temperature difference and molecular weight of the atmosphere. Importantly, this implies that the wind speeds become independent of molar heat capacity and the mean temperature of the atmosphere. The high-molecular-mass limit—which corresponds to the limit where the day-night temperature difference converges toward radiative equilibrium—only applies when the wave timescale is sufficiently long compared to the radiative timescale. Nevertheless, this condition readily becomes satisfied when the molecular mass is large, $\mu > 10$ g mol$^{-1}$ at low pressures in our simulations, for example. Once that limit is reached, the wave and radiative timescales no longer directly influence the wind speeds. Our simulations seem to support this argument. The wind speeds exhibit the same trend in Experiments II and III in which the radiative timescale is fixed for all cases, implying that the radiation timescale does not control the wind speed. We also did an extra experiment (not shown here) based on the standard cases with mean temperature enhanced by a factor of 3 but keeping $\Delta T_{eq}$. The resulting $U_{rms}$ is roughly the same as that in Experiment I, implying that the mean thermal state has negligible effect on the wind speed. Finally, the agreement of Experiments I and II suggests that the molar heat capacity effect is not important (Fig. 5).

### 4.4. Trend 4: Jet Width

In the zonal-mean zonal flow pattern (Fig. 3 and 4), the number and width of the zonal jets depends on the molecular mass. Hydrogen atmospheres exhibit only one broad jet—an eastward jet centered at the equator—while higher molecular-mass atmospheres have a narrower equatorial jet, flanked by regions of westward flow at higher latitudes. At the highest values of molecular weight (e.g., $N_2$ and $CO_2$), eastward zonal-mean zonal wind emerges close to the poles. The latitudinally alternating structure in the zonal-mean zonal wind map of the $CO_2$ case is consistent with the simulations by Kataria et al. (2014) with a fully realistic radiative transfer for GJ1214b, although their polar jets have higher speeds.

The half-width of the equatorial jet can be estimated by the equatorial deformation radius $L_e \sim (NH/\beta)^{1/2}$ that measures a decaying length scale of long equatorial Rossby waves (Showman & Polvani 2011). We assume an isothermal temperature profile and use Eq. (4) to estimate the jet width:

$$L_J \sim 2L_e \sim 2\left(\frac{TR^2}{\mu c_p \beta^2}\right)^{1/4}. \qquad (25)$$

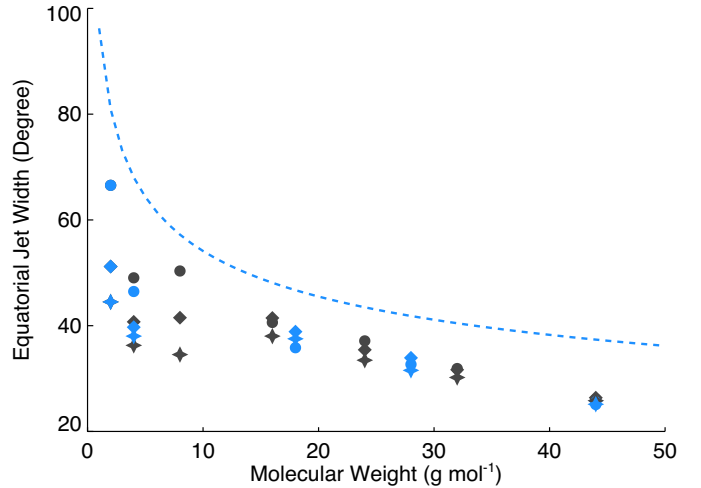

FIG. 13.— The e-folding widths of equatorial jets versus molecular weight at 700 Pa (dots), 5700 Pa (diamonds) and 23100 Pa (stars) for Experiment II. Blue line is prediction of Eq. (25).

The predicted wind speeds generally follow the trend of the simulations but are slightly higher than the e-folding width of the simulated jets (Fig. 13). Our estimate suggests the jet width does not significantly vary with pressure, which seems consistent with the simulations, especially in the high molecular weight regime (Fig. 13).

### 4.5. Trend 5: Pressure of the Equatorial Jet Core

A strong radiative relaxation acts to suppress the jets formation and damp the jets (Showman et al. 2013). These authors showed that, if the radiative time constant is short, Rossby and Kelvin waves, which cause the phase tilts and drive the equatorial jet formation in a synchronously rotating atmosphere, will be suppressed—as a result, the equatorial superrotating jet is damped. Showman et al. (2013) argued that the suppression of jet formation should occur when the radiative time constant is shorter than the timescale for the Kelvin wave to propagate over a hemisphere, i.e., $\tau_{rad} < \tau_{wave}$. In this case, the atmosphere exhibits a day-to-night flow pattern. If the radiative timescale is longer than the wave propagation timescale ($\tau_{rad} > \tau_{wave}$), the atmosphere is dominated by an equatorial jet. In our case, the radiative timescale is small in the upper atmosphere and increases with pressure (e.g., Fig. 2), and therefore it is possible that the general circulation pattern shifts from a day-to-night divergent flow in the upper atmosphere to a jet-dominated flow in the deeper atmosphere (Showman et al. 2013). This is consistent with our simulations where the equatorial zonal jet only occurs below some pressure level (Fig. 3).

This transition from a zonal flow at depth to a divergent day-to-night flow aloft suggests that the zonal-mean zonal wind maximizes at some specific pressure near this transition. In the regime of zonal jets deep in the atmosphere, the zonal wind has similar sign at many longitudes along a given latitude circle, and thus the zonal-mean zonal wind is strong. Our models suggest that at and below the photosphere level, the zonal-mean temperature is generally warmer at the equator than at high latitudes, and this implies through a generalized thermal-wind balance that, starting from some deep level of assumed weak winds, the zonal-mean zonal winds will increase with height in the jet-dominated regime. However, in the regime of day-to-night flow, there is a significant cancellation of the eastward- and westward-flowing branches along any given latitude circle, which causes the zonal-mean zonal wind to become weak, even if the local values of the day-night eddy winds themselves are large. Thus, in the divergent flow regime, one might expect the zonal-mean zonal winds to decrease with increasing altitude. The implication is a maximum of zonal-mean zonal wind with pressure near the transition between these regimes. Interestingly, at altitudes above this maximum, the decrease of zonal-mean zonal winds with altitude would necessarily imply a reversal of the zonal-mean meridional temperature gradient (with the poles being warmer than the equator in the zonal mean). Understanding the details of this phenomenon are beyond the scope of the present paper, although we note that the phenomenon is commonly associated with wave-driven circulations that are common in planetary stratospheres throughout the solar system. Our simulations are qualitatively consistent with both a regime transition from jet to divergent flow with increasing altitude and likewise generally exhibit a maximum zonal-mean zonal wind which occurs at approximately similar pressure to the transition (see Figs. 3-4), suggesting validity to our overall argument.

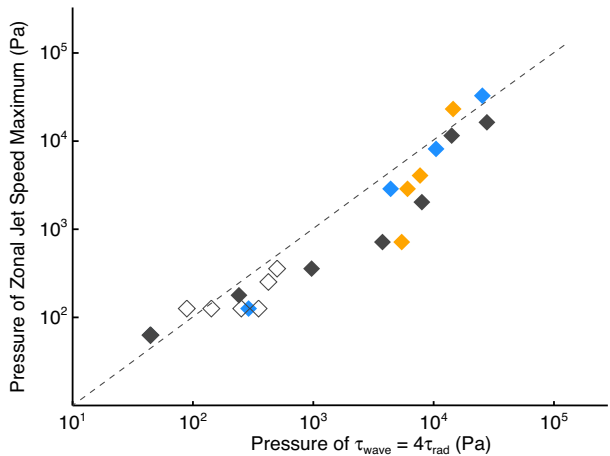

Fig. 14.— Correlation between the zonal jet core pressure and the pressure level where $4\tau_{rad} = \tau_{wave}$, with a unity-slope dashed line. Experiments I to IV in Table 2 corresponds to symbols of blue, black filled, black open, orange filled, respectively. $\tau_{wave}$ is zonal averaged for all experiments.

We expect the jet speed maximum occurs at the pressure level where the radiative timescale is roughly equal to the wave propagation timescale. Empirically, the pressure levels at which the equatorial zonal-mean jet speed maximizes ("the jet core pressure") in our simulations correlate well with which the radiative timescale is equal to a quarter of the gravity wave propagation timescale (Fig. 14), i.e., $\tau_{wave} \sim 4\tau_{rad}$. Based on the approximations of $\tau_{wave}$ (Eq. 3) and $\tau_{rad}$ (Eq. 7 and 8) in Section 2, using $L \sim R_p$, we obtain the analytical expression of jet core pressure $p_J$:

$$p_J \sim \left(\frac{R_p}{4KRT^{1/2}}\right)^{4/3} \frac{\mu^2}{c_p^{2/3}}. \qquad (26)$$

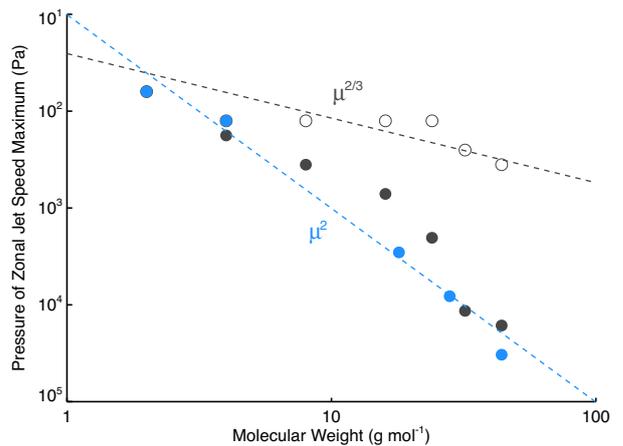

Fig. 15.— Jet core pressure as a function of molecular weight for Experiments I (blue filled), II (black filled) and III (open circles). The blue and black dashed lines represent the analytical expressions from Eq. (26) and (27), respectively.

Note that this expression is related to the prescribed formulation of radiative timescale in this study $\tau_{rad}(p) = Kp^{3/4}\mu^{-1}c_p$ (Eq. 8), where $K$ is a constant [2]. However, it is straightforward to generalize the above analytical expression to a more realistic case, given the pressure dependence of radiative timescale in the real atmosphere. Eq. (26) predicts that the atmospheric zonal jet occurs deeper in the atmosphere as molecular weight increases, consistent with Figs. 3 and 4. Quantitatively, if the molar heat capacity is fixed (Experiment II), the jet core pressure $p_J$ scales as $p_J \propto \mu^2$ from Eq. (26). This analytical expression matches our simulations quite well (Fig. 15).

The radiation effect has a significant impact on the jet core pressure level. The simulations with the radiative timescale fixed (black open circles in Fig. 15) exhibit a jet-core peak at much lower pressures than their control cases (black dots). If radiative timescale is taken to be independent of molar heat capacity and molecular mass, the jet core pressure $p_J$ is:

$$p_J \sim \left(\frac{\mu_0 R_p}{4c_{p0}KRT^{1/2}}\right)^{4/3} \mu^{2/3}c_p^{2/3} \qquad (27)$$

where $c_{p0}$ and $\mu_0$ are the molar heat capacity and molecular weight for molecular hydrogen, respectively. This analytical expression agrees well with our simulations (Fig. 15). In this case, the jet core pressure $p_J$ scales as $p_J \propto \mu^{2/3}$ from Eq. (27).

### 4.6. Trend 6: Maximum Zonal-mean Jet Speed $U_z(p_J)$

The generation of equatorial prograde jets has been studied for decades but this problem has not been com-pletely solved yet. The equatorial jet formation on tidally locked exoplanets might be closely related to the standing Rossby wave pattern (Showman & Polvani 2011) and the resonance of the Rossby waves (Tsai et al. 2014). Normally the zonal-mean jet speed is difficult to estimate without knowing the details of the wave-mean flow interactions (Vallis 2006). No theory currently exists that can quantitatively predict the zonal-mean zonal wind speed on synchronously rotating planets. Nevertheless, the Komacek & Showman (2016) theory, and our alternate form of that theory presented in Section 4.3 and Appendix A, does provide a theoretical prediction for the characteristic horizontal wind speed $U_{rms}$. In the maximum zonal jet speed region, the contribution of zonal velocity $u$ is much larger than the meridional velocity $v$ at the equator, and $U_{rms}$ (Eq. 11) in the equatorial region can be approximated by:

$$U_{rms}(p_J) \approx \left( \int_{-\pi/6}^{\pi/6} \overline{u(\lambda, \phi, p_J)^2} \cos \phi d\phi \right)^{1/2} \quad (28)$$

where $p_J$ denotes to the jet core pressure (Eq. 26). As noted in Section 3, $U_{rms}(p_J)$ is related to the equatorial zonal-mean zonal wind ($U_z$, Eq. 12) and the eddy component ($U_e$, Eq. 13) via $U_{rms}(p_J)^2 \approx U_z(p_J)^2 + U_e(p_J)^2$. Furthermore, in the jet core region, we found that the $U_z(p_J)$ is significantly larger than $U_e(p_J)$. In fact, the zonal-mean kinetic energy occupies about 95% of the total kinetic energy in the jet core region, i.e., $U_z(p_J)^2/U_{rms}(p_J)^2 \sim 95\%$. Therefore, we can approximate the maximum zonal-mean jet speed (the "jet core speed") $U_z(p_J)$ using $U_{rms}(p_J)$. Note that this approximation only works in the jet core region where the kinetic energy is primarily dominated by the zonal-mean component.

---

[2] From Eqs. (7) and (8), one can derive the constant $K = 10^{5/2}(4/7R)$.

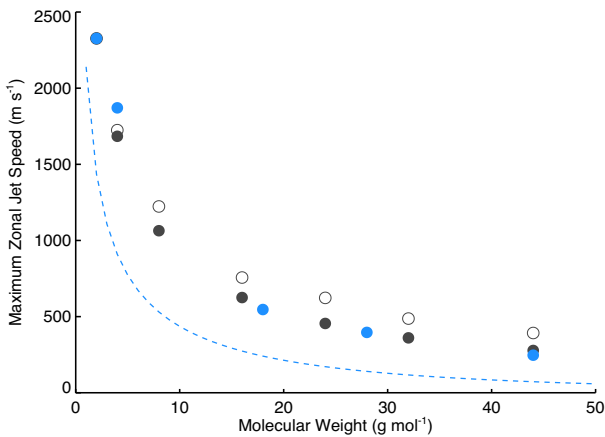

FIG. 16.— Maximum zonal jet speed versus molecular weight for the standard cases (blue filled), Experiment II (black filled) and III (open circles). For comparison, the dashed line is the theoretical prediction of the characteristic RMS wind from the theory of Komacek & Showman (2016) and as modified here (Equation 22 and Appendix A), evaluated at the pressure of the jet-core peak predicted by Equation (24). Note that a formal theory for the zonal-mean zonal wind on synchronously rotating exoplanets does not yet exist.

$U_{rms}(p_J)$ should decrease with molecular weight because (1) at the same pressure level, $U_{rms}$ decreases with $\mu$ (Fig. 11); and (2) the pressure of the equatorial jet core $p_J$ increases with $\mu$ (Fig. 15) but $U_{rms}$ decreases with increasing pressure (Fig. 12). Using Eq. (22) and (24), we can analytically predict $U_{rms}(p_J)$ and thus the maximum zonal-mean zonal wind $U_z(p_J)$. For hydrogen atmosphere, $U_z(p_J)$ is on the order of 1000 m s$^{-1}$ and that for CO$_2$ atmosphere is on the order of 100 m s$^{-1}$. The predicted equatorial jet core speeds as a function of molecular weight are approximately consistent with simulations within a factor of 2 (Fig. 16).

## 5. EFFECTS OF MOLAR HEAT CAPACITY

The effect of heat capacity can be inferred from the difference between the standard cases (Experiment I) and the cases with heat capacity fixed (Experiment II). Because the variation of the molar heat capacity around 1000 K is generally within a factor of three (from $c_p \sim 2.5R$ for helium to $c_p \sim 6.5R$ for carbon dioxide, see Fig. 1), the effect is generally limited. Furthermore, the two important parameters in our analytical theory in Appendix A, $\alpha$ and $\gamma$, are a function of $\tau_{rad}/\tau_{wave}^2$. In an isothermal atmosphere, $\tau_{rad} \propto c_p$ (Eq. 8) and $\tau_{wave} \propto c_p^{1/2}$ (Eq. 3), implying that $\tau_{rad}/\tau_{wave}^2$ does not depend on $c_p$. As a result, both day-night temperature contrast and global wind speed are not strongly influenced by the molar heat capacity, as seen in Fig. 7 and 10, respectively.

However, the molar heat capacity seems to have effects on the zonal-mean zonal wind pattern in our simulations, as seen in Experiment IV based on the "mass-16" case with different heat capacity (Fig. 17). It shows that the zonal-mean equatorial jet speed decreases with increasing heat capacity, but the variation is within a factor of 2. The most significant impact is the jet core pressure. Cases in Experiment IV exhibit a dramatic change of the jet core pressure while the molar heat capacity only varies by a factor of 2 (from $7R/2$ to $14R/2$). The circulation pattern at 700 Pa (Fig. 17) transits from the equatorial-jet-dominated regime in the $7R/2$ case to a day-to-night flow regime in the $14R/2$ case.

That the jet core pressure moves deeper into the atmosphere in the simulations contradicts our theory of jet core pressure (Eq. 26) in Section 4.5, which predicts that the jet core pressure $p_J$ should scale with $c_p^{-2/3}$ and thus be smaller for a higher $c_p$ atmosphere. As $c_p$ increases, the radiative timescale ($\tau_{rad} \propto c_p$, Eq. 8) increases more than the wave timescale ($\tau_{wave} \propto c_p^{1/2}$, Eq. 3) in an isothermal atmosphere, implying a less radiative-control atmosphere and the zonal jet should form at a lower pressure level. Furthermore, $p_J \propto c_p^{-2/3}$ is a very weak dependence given the variation of $c_p$ within a factor of 2. Neither the sign nor the magnitude of the $c_p$ dependence in Fig. 17 are consistent with Eq. (26), albeit which has successfully explained the principal trend of the regime shift and its dependence on the molecular weight (Fig. 15).

What leads to the discrepancy between Experiment IV and our theory in Section 4.5? We hypothesize that it is the assumption of an isothermal atmosphere used for wave propagation timescale (Eq. 3). A more realistic estimate of $\tau_{wave}$ should be based on the $NH$ given in Eq. (2). If the atmospheric temperature is close to adiabatic, a small change of the molar heat capacity might greatly reduce the gravity wave speed $NH$ in the atmosphere and inhibit equatorial jet formation. In this situation, the dependence of jet core pressure on $c_p$ could be significant, as in Experiment IV.

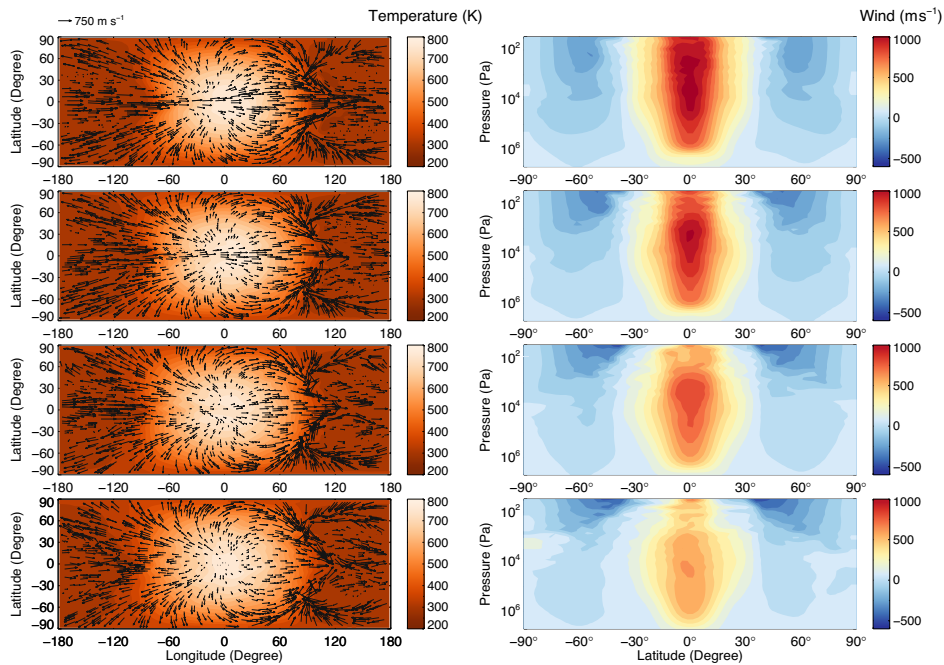

FIG. 17.— Longitude-latitude temperature and wind maps at 700 Pa (left column) and the zonal-mean zonal wind ($\overline{u}$) distributions (right column) as a function of pressure and latitude from Experiment IV, the cases with different molar heat capacity but the same molecular weight ($\mu$=16 g mol⁻¹). From top to bottom: the $c_p$ of $7R/2$, $10R/2$, $12R/2$ and $14R/2$, respectively.

We separately analyzed the wave propagation timescales on the dayside and the nightside, and this analysis suggests that the planetary-scale Kelvin/Rossby waves are more easily suppressed on the nightside than the dayside. Using the local temperature and its gradient with pressure, we locally calculated $NH$ at each (longitude, latitude) point over the globe. For each pressure level, the *mean* wave propagation timescales for *each hemisphere* is approximated by $\tau_{wave} = R_p/\overline{NH}$ where $\overline{NH}$ is obtained by spatially averaging the $NH$ values for each hemisphere. These averages include only points within 30° latitude of the equator. Fig. (18) shows that $\tau_{wave}$ decreases on both the dayside and the nightside as the molar heat capacity increases, but the change of $\tau_{wave}$ is much more dramatic on the nightside (more than an order of magnitude) than on the dayside (less than 50%) [3]. This is because the nightside hemisphere is less vertically stratified than the dayside—the nightside atmosphere has a larger vertical temperature gradient

($\partial \ln T/\partial \ln p$) [4]. Based on Eq. (2), the wave propagation speed is smaller in the nightside. The fractional change of $\overline{NH}$ due to $c_p$ increase could also be larger on the nightside. The quantity $NH$ is more sensitive on the nightside than on the dayside to variations in $c_p$, allowing $c_p$ to exert a greater influence on wave dynamics on the nightside than on the dayside. This sensitivity results simply from the fact that the thermal profile $\partial \ln T/\partial \ln p$ is closer to an adiabat on the nightside than on the dayside, especially in the $c_p = 14R/2$ case (Fig. 18).

Invoking our previous theory that the jet core pressure correlates with the pressure where the radiative timescale is roughly equal to a quarter of the wave propagation timescale in Section 4.2, we expect the jet core pressure to be given by the pressure where the vertical profiles of $4\tau_{rad}$ and $\tau_{wave}$ intersect. The radiative timescales are plotted on top of $\tau_{wave}$ as a function of pressure in Fig. 18. Interestingly, the trend of the crossing points as function of the molar heat capacity shows an opposite behavior between the dayside and the nightside. On the dayside, the crossing point moves towards the upper atmosphere as the molar heat capacity increases. On the other hand, on the nightside, the crossing point moves deeper into the atmosphere, consistent with the jet core pressure trend in Experiment IV, implying that nightside wave dynamics might be more important in controlling the transition from the day-to-night flow to the jet-dominated flow when the atmospheric mean temper-

---

[3] Here we assume the wave speed is a vertically *local* quantity that is equal to $NH$. This is based on the assumption that $NH$ is constant with height. If $NH$ varies with height (as in Fig. 18, nightside), then the appropriate wave speed would involve a an appropriate vertical integral of $N$ with height over the range of heights associated with that wave mode. Therefore the extremely short wave speeds (and corresponding long wave timescales) at $10^3$–$10^4$ Pa in Fig. 18 (orange dashed curve) might not be realistic (although the wave speed could still be somewhat short), because the wave speed might be significantly influenced by adjacent layers that are more stratified.

[4] The difference of vertical temperature gradient between dayside and nightside can be expressed as:

$$\frac{\partial \ln T_{day}}{\partial \ln p} - \frac{\partial \ln T_{night}}{\partial \ln p} = \frac{\partial (\ln T_{day} - \ln T_{night})}{\partial \ln p} < 0. \quad (29)$$

The inequity originates from that the day-night temperature contrast ($T_{day} - T_{night}$) decreases with increasing pressure (Section 4.1).

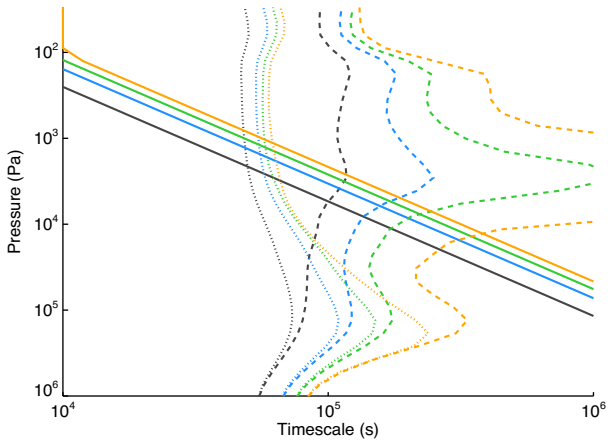

Fig. 18.— Timescales of radiative relaxation ($4\tau_{rad}$, solid), dayside gravity wave propagation (dotted) and nightside gravity wave propagation (dashed), as a function of pressure. The colors represent four molar heat capacity cases: $7R/2$ (black), $10R/2$ (blue), $12R/2$ (green), and $14R/2$ (orange). From low to high heat capacity cases, the crossing points between the radiative timescale and wave propagation timescale moves towards upper atmosphere in the dayside but towards deeper atmosphere in the nightside.

ature is close to adiabatic.

## 6. CONCLUDING REMARKS

Sub-Jupiter size planets dominate the population of currently known planets. Due to the limitation of the observational techniques, we expect most discovered planets will be close to their host stars. This emerging population shall exhibit a larger-than-ever variety of chemical compositions in their atmospheres. Here we took a preliminary step to study the influence of their bulk composition on the planetary-scale atmospheric circulation and the temperature distributions. Future transit measurements, possible direct imaging observations and Doppler imaging techniques will provide a vast amount of information on their atmospheres and may test the systematic behaviors predicted in this study.

We have mainly investigated the influences on atmospheric dynamics from two independent properties of the bulk composition: molecular weight and molar heat capacity. As the molecular weight increases, the atmosphere tends to have a larger day-night temperature contrast, a smaller RMS wind speed, a smaller eastward phase shift in the thermal phase curve, and a narrower equatorial super-rotating jet that moves to deeper into the atmosphere. The zonal-mean zonal wind is smaller and more likely to exhibit a latitudinally alternating pattern in a higher-molecular-weight atmosphere. We also found if the vertical temperature profile is close to adiabatic, the molar heat capacity will play a significant role in controlling the transition from a divergent flow in the upper atmosphere to a jet-dominated flow in the lower atmosphere.

The underlying physical mechanisms have been discussed both qualitatively and quantitatively. The influences of molecular weight and molar heat capacity mainly originate from the interplay between the radiation and wave dynamics. The molecular weight also has a direct impact on the wind speed. We summarize our diagnosis as below:

(1) Most of the trends in our simulations of possible bulk compositions are due to the change of molecular weight, mainly because, for plausible compositions, molecular weight varies over a much wider range than the molar heat capacity.

(2) The temperature contrast between the dayside and nightside increases with the molecular weight. As the molecular weight increases, the atmosphere exhibits a temperature structure that becomes closer to radiative equilibrium. Perez-Becker & Showman (2013) and Komacek & Showman (2016) showed that the day-night temperature difference is controlled by a competition between the tendency of zonally propagating planetary-scale waves to mute the day-night temperature difference, and the tendency of the day-night radiative-heating contrast to increase it. The wave speeds decrease with increasing molecular weight, which lessens the ability of the waves to adjust the thermal structure. Therefore, this wave-adjustment mechanism becomes more easily suppressed–leading to larger day-night temperature differences–at high molecular weight.

(3) The wind speeds decrease with increasing molecular weight. This results from the fact that large molecular weight implies small scale height, which implies a smaller day-night difference in the geopotential (on isobars) than would occur in a low-molecular-mass atmosphere. This causes a weaker pressure-gradient force in the horizontal momentum equation, and therefore leads to weaker winds.

(4) The peak of thermal phase curve is shifted prior to the secondary eclipse due to a strong superrotating jet at equator. The phase shift is controlled by the ratio of atmospheric radiative timescale versus advection timescale of the zonal-mean zonal wind. The phase shift is smaller if the molecular weight is larger because the radiative timescale decreases and advection timescale increases with the molecular weight.

(5) The pressure level at which the zonal-mean jet speeds reach a maximum roughly coincides with the pressure level where the radiative timescale is equal to a quarter of the wave propagation timescale in our simulations, implying that the jet core pressure is strongly controlled by both radiative effects and dynamical effects. In the jet core region, the maximum zonal-mean jet speed $U_z(p_J)$ can be approximately by $U_{rms}(p_J)$, which decreases with molecular weight.

(6) As molecular weight increases, the equatorial jet width becomes narrower. More alternating wind patterns emerge in the higher latitude in the zonal-mean zonal wind distribution. The $CO_2$ atmosphere exhibits five alternating wind patterns in the zonal-mean zonal wind map. Showman & Polvani (2011) predicted that the half-width of the equatorial-jet scales with the equatorial deformation radius. We find that this theory explains the jet widths in our simulations; when the molecular mass is greater, the equatorial deformation radius is smaller, and the equatorial jet is narrower.

(7) Consistent with previous work (Showman et al. 2013), the longitude-latitude wind pattern shifts from a day-to-night divergent flow pattern in the upper atmosphere to the jet-dominated pattern in the lower atmosphere. As molecular weight increases, the transition occurs deeper in the atmosphere.

(8) The effect of molar heat capacity is generally small except when the atmospheric temperature profile is close to adiabatic, which occurs most easily on the nightside.

The effect becomes important primarily when the molar heat capacity is large. In this situation, a slight increase of the molar heat capacity might greatly reduce wave speed in the atmosphere, particularly in the nightside atmosphere, and significantly influence the jet formation in the upper observable atmosphere. As a result, the pressure of the transition from a day-night divergent flow to a jet-dominated flow moves deeper into the atmosphere.

Several analytical theories have been presented in this study. We presented a modified form of the theory from Komacek & Showman (2016) that yields predictions of the day-night temperature difference and root-mean-square wind speeds $U_{rms}$ as a function of radiative time constant, frictional drag time constant, rotation rate, and other parameters. This theory explains well the dependence of day-night temperature difference and wind speed with molecular weight. We presented an analytic estimate of the eastward offset of the dayside hot spot if the equatorial jet speed is known. We showed that the equatorial jet width is consistent with the equatorial deformation radius. We also estimated the pressure ($p_J$) at which the zonal-mean equatorial jet reaches a maximum. We furthermore showed that the analytic estimate of $U_{rms}$ provides a reasonable estimate to the dependence of the peak zonal-*mean* zonal wind $U_z(p_J)$ with molecular weight. This should provide constraints for future theories of what controls the equilibrated speed of the zonal-mean zonal wind, which have yet to be constructed. All of the above estimates are closely related to possible observables on tidally locked exoplanets.

In terms of observational implications, aside from the trends that were summarized above and might be tested in statistics with a large sample of data, we also emphasize two points here, related to the hazes and clouds which are recently identified on several super Earths.

First, we found a correlation between the atmospheric molecular weight and the jet speed and the width of alternating zonal-mean wind pattern in the atmosphere. That might shed light on a new way to infer the atmospheric composition from the wind pattern observations which might be achieved from a precise Doppler imaging measurement (e.g., Snellen et al. 2010). This method might be typically useful for the planetary atmospheres covered by the hazes and clouds, which flatten the spectrum and make it difficult to infer broad spectral features diagnostic of the composition, as in the case of GJ1214b (Kreidberg et al. 2014). The general circulation simulations with more realistic radiative transfer scheme from (Kataria et al. 2014) also show that the jet pattern is significantly distinct between the water atmosphere and the carbon dioxide atmosphere.

Second, if the atmospheric radiative equilibrium situation is close to adiabatic, as seen in some high molar heat capacity simulations, the nightside atmosphere has a large tendency to trigger convection above the zonal jet region once the temperature gradient drops so that the Richardson number $Ri \equiv N^2/(\partial^2 \overline{u}/\partial z^2)$ is under 0.25 (Chandrasekhar 1961). In general, haze and cloud particles might be preferentially formed in the nightside of a colder environment. If the convection occurs in the nightside atmosphere, it will enhance the vertical mixing of those particles in the atmosphere (Parmentier et al. 2013; Lee et al. 2015; Charnay et al. 2015a). It might help to loft the haze/cloud particles to the upper atmosphere and stabilize the low-pressure haze layers observed on several super Earths/mini Neptunes (e.g., Kreidberg et al. 2014; Knutson et al. 2014a; Knutson et al. 2014b; Ehrenreich et al. 2014).

Finally, it should be noted that this work is highly idealized with a simple Newtonian cooling radiative scheme in order to highlight the underlying physical mechanisms. Our analytical theories are also based on the Newtonian cooling assumption. With a gray radiative transfer scheme, a recent work by Komacek et al. (2016) has shown that the theories of day-night temperature contrast and thermal phase shift presented here could also explain their simulations for hot Jupiters, implying that those theories might be applicable to more realistic situations. The opacity issue, although crucially important, is very complicated and beyond the scope of this paper. To the first order, different opacity corresponds to different pressure level where $\tau = 1$ in Eq. (6) and implies different pressure dependence of the radiative timescale. The quantitative prediction related to the radiative timescale would thereby change accordingly. But the underlying mechanisms discussed in this study remain the same. We also did not explore the other parameters such as the situation of non-synchronized rotating planets (Showman et al. 2015), eccentric orbits (Lewis et al. 2010; Lewis et al. 2013; Kataria et al. 2013), or planets with non-zero obliquities. In those situations, a whole range of atmospheric behaviors will emerge and the dynamics in a larger parameter space is far more rich than what we have discussed here.

## APPENDIX

### APPENDIX A: A UNIFORM EXPRESSION FOR DAY-NIGHT TEMPERATURE DIFFERENCE AND WIND SPEED ON TIDALLY LOCKED PLANETS

As shown in Perez-Becker & Showman (2013) in shallow water equations and Komacek & Showman (2016) in the three-dimensional primitive equations, we can obtain analytical expressions for a proxy of day-night temperature difference ($\Delta T/\Delta T_{eq}$) and horizontal wind speed $U$, which should be considered as the RMS wind velocity $U_{rms}$, on tidally locked planets. The horizontal pressure-gradient force associated with the horizontal day-night temperature difference drives the flow. This force will be balanced by the largest of the remaining forces, namely, frictional drag, Coriolis force, horizontal momentum advection, and vertical momentum advection (Komacek & Showman 2016). This leads to distinct expressions for the day-night temperature difference and wind speed in each regime. Here, we construct a modified version of this theory in which the four regimes are combined into a single expression that naturally transitions from one regime to another as appropriate.

We adopted the same assumptions as in Komacek & Showman (2016) to simplify the primitive equations. We assume that the day-night temperature difference extends vertically over a log pressure range of $\Delta \ln p = \ln(p_{bot}/p)$ where $p_{bot}$ is the pressure level deep in the atmosphere at which the temperature is horizontally homogeneous. The magnitude of the horizontal acceleration term in the momentum equation (Eq. 1a) can be estimated as $R\Delta T \Delta \ln p/2\mu L$ (Komacek & Showman 2016). For the thermodynamical equation (Eq. 1d), following Komacek & Showman (2016), we adopted the "Weak-Temperature Gradient" (WTG) approximation (Sobel et al. 2001), in which the horizontal advection of the temperature is neglected. Assume $f \sim \Omega$ and the scaling equation set is:

$$\frac{U^2}{L} + \Omega U \sim \frac{R\Delta T \Delta \ln p}{2\mu L} - \frac{U}{\tau_{drag}} \tag{A.1a}$$

$$\frac{\Delta T_{eq} - \Delta T}{2\tau_{rad}} \sim \frac{wN^2 H\mu}{R} \tag{A.1b}$$

$$\frac{U}{L} \sim \frac{w}{H} \tag{A.1c}$$

See Komacek & Showman (2016) for the derivation of Eq. (A. 1a) and Eq. (A. 1b). Here we retained a factor of 2 in both equations for a more quantitative theory-simulation comparison. For simplicity, Komacek & Showman (2016) dropped a factor of 2 in their Eq. (27) when the geopotential height was integrated with pressure (see their Eq. 25). Here we retain that factor in the right hand side of Eq. (A.1a). Based on the Fig. 6 of Komacek & Showman (2016), we revised their Eq. (22) as $\Delta T_{eq} - \Delta T \sim 2|T_{eq} - T|_{global}$, leading to the factor of 2 in the left hand side of Eq. (A.1b).

We should emphasize that here we do not intend to solve the primitive equations (Eq. 1). The formulation of Eq. (A.1) is only for a purpose of combining the possible dominant terms: advection, rotation and drag, to achieve a uniform expression for the same theory that has been derived in Komacek & Showman (2016). Hopefully the uniform expression is more general and might apply to the transitions between those four limiting regimes. Combine the above three equations yields the solutions for $\Delta T$ and $U$:

$$\frac{\Delta T}{\Delta T_{eq}} \sim 1 - \frac{2}{\alpha + \sqrt{\alpha^2 + 4\gamma^2}} \tag{A.2}$$

$$\frac{U}{U_{eq}} \sim \frac{2\gamma}{\alpha + \sqrt{\alpha^2 + 4\gamma^2}} \tag{A.3}$$

where $U_{eq} = (R\Delta T_{eq} \Delta \ln p/2\mu)^{1/2}$ can be considered as the cyclostrophic wind speed induced by an equilibrium day-night temperature difference. The non-dimensional parameters $\alpha$ and $\gamma$ are defined as:

$$\alpha = 1 + \frac{(\Omega + \frac{1}{\tau_{drag}})\tau_{wave}^2}{\tau_{rad}\Delta \ln p} \tag{A.4}$$

$$\gamma = \frac{\tau_{wave}^2}{\tau_{rad}\tau_{adv,eq}\Delta \ln p} \tag{A.5}$$

where $\tau_{wave} \sim L/NH$ is the timescale for wave propagation. $\tau_{adv,eq} \sim L/U_{eq}$ is the advective timescale due to the "equilibrium cyclostrophic wind" $U_{eq}$. Note that Eq. (A.3) can also be written as:

$$\frac{U}{U_{eq}} \sim \sqrt{(\alpha/2\gamma)^2 + 1} - \alpha/2\gamma \tag{A.6}$$

Therefore, the ratio of the wind speed to the "equilibrium cyclostrophic wind" $U_{eq}$ solely depends on a single parameter $\alpha/2\gamma$.

The parameter $\alpha$ is related to the Coriolis term which is characterized by the timescale of $\Omega^{-1}$, and the drag term which is characterized by the timescale of $\tau_{drag}$. $\gamma$ is related to the advection term characterized by the timescale of $\tau_{adv,eq}$. By comparing the magnitudes of the three timescales, we can obtain the limits of Eq. (A.4) and (A.5).

(1) When drag dominates, $\tau_{drag}$ is much smaller than $\Omega^{-1}$ and $\tau_{adv,eq}$, and the parameter $\alpha$ is reduced to $\alpha_d$:

$$\alpha_d = 1 + \frac{\tau_{wave}^2}{\tau_{drag}\tau_{rad}\Delta \ln p}. \tag{A.7}$$

(2) In the Coriolis-dominated regime, rotation dominates, and therefore $\Omega^{-1}$ is much smaller than $\tau_{drag}$ and $\tau_{adv,eq}$. The parameter $\alpha$ is reduced to $\alpha_i$ in the inviscid limit:

$$\alpha_i = 1 + \frac{\Omega\tau_{wave}^2}{\tau_{rad}\Delta \ln p}. \tag{A.8}$$

(3) In the advection-dominated regime, $\tau_{adv,eq}$ is much smaller than $\Omega^{-1}$ and $\tau_{drag}$, which is equivalent to $\alpha \sim 1$.

In the above three regimes, the day-night temperature difference is:

$$\frac{\Delta T}{\Delta T_{eq}} \sim \begin{cases} 1 - \frac{1}{\alpha_d} & \text{drag dominated} \\ 1 - \frac{1}{\alpha_i} & \text{Coriolis dominated} \\ 1 - \frac{2}{1+\sqrt{1+4\gamma^2}} & \text{advection dominated} \end{cases} \tag{A.9}$$

Similarly, for the wind speed $U$, we have corresponding limits:

$$\frac{U}{U_{eq}} \sim \begin{cases} \frac{\gamma}{\alpha_d} & \text{drag dominated} \\ \frac{\gamma}{\alpha_i} & \text{Coriolis dominated} \\ \frac{2\gamma}{1+\sqrt{1+4\gamma^2}} & \text{advection dominated} \end{cases} \tag{A.10}$$

One can show that the above expressions for day-night temperature difference are consistent with the ones derived in Komacek & Showman (2016). We emphasize that the characteristic wind speed $U$ we derived here should not be confused with the zonal-mean zonal wind $\overline{u}$, and should instead be considered as the RMS wind velocity $U_{rms}$, although in the jet core region, the two values might be close (Section 4.6).

APPENDIX B: A KINEMATIC MODEL OF TEMPERATURE DISTRIBUTION AND THERMAL PHASE SHIFT

### B.1. Temperature Distribution

We construct a simple kinematic model for the thermal phase shift in the presence of a specified zonal jet. For simplicity, we first consider an 1D solution. In the zonal direction, the thermodynamic equation can be reduced to a 1D equation with respect to longitude $\lambda$. We adopt the Newtonian cooling approximation for the radiative scheme by assuming a constant radiative timescale $\tau_{rad}$ and a constant "advective" timescale $\tau_{adv} \sim L/\overline{u}$ where $\overline{u}$ is the zonal-mean zonal wind. The kinematic equation is:

$$\frac{\partial T}{\partial t} + \frac{1}{\tau_{adv}}\frac{\partial T}{\partial \lambda} = \frac{T_{eq}(\lambda) - T}{\tau_{rad}} \tag{B.1}$$

where $T_{eq}$ is the equilibrium temperature, the distribution of which could be given by Eq. (9) for a tidally locked planet. Since we have decoupled the latitude and altitude with longitude in this 1D problem, we assume:

$$T_{eq}(\lambda) = \begin{cases} T_n + T_1 \cos\lambda & \text{dayside} \\ T_n & \text{nightside} \end{cases} \tag{B.2}$$

where we assume the equilibrium temperature in the nightside is flat: $T_n = T_0 - T_1/2$. $T_0$ is the global mean temperature and $T_1$ is the temperature difference between the sub-stellar and the anti-stellar points. Here we define $\lambda = 0$ as the longitude of the substellar point and solve Eq. (B.1) in steady state in the domain $[-\pi, \pi]$:

$$\xi\frac{\partial T}{\partial \lambda} = \begin{cases} T_n + T_1 \cos\lambda - T & \text{dayside } (-\pi/2 \leq \lambda \leq \pi/2) \\ T_n - T & \text{nightside } (-\pi \leq \lambda \leq -\pi/2 \ \& \ \pi/2 \leq \lambda \leq \pi) \end{cases} \tag{B.3}$$

where $\xi = \tau_{rad}/\tau_{adv}$ is the ratio between the radiative timescale and the advective timescale. The general solution is:

$$T(\lambda) = \begin{cases} T_n + T_1 \cos\lambda_s \cos(\lambda - \lambda_s) + k_1 e^{-\lambda/\xi} & -\pi/2 \leq \lambda \leq \pi/2 \\ T_n + k_2 e^{-\lambda/\xi} & -\pi \leq \lambda \leq -\pi/2 \\ T_n + k_3 e^{-\lambda/\xi} & \pi/2 \leq \lambda \leq \pi \end{cases} \tag{B.4}$$

where $\lambda_s = \tan^{-1}\xi$. $k_1$, $k_2$ and $k_3$ are determined by the continuity of solutions at terminators $T(\pm\pi/2)$ and periodic boundary condition $T(-\pi) = T(\pi)$. The final solution of Eq. (B.1) can be expressed as:

$$T(\lambda) = \begin{cases} T_n + T_1 \cos\lambda_s \cos(\lambda - \lambda_s) + \eta T_1 e^{-\lambda/\xi} & -\pi/2 \leq \lambda \leq \pi/2 \\ T_n + \eta T_1 e^{-(\pi+\lambda)/\xi} & -\pi \leq \lambda \leq -\pi/2 \\ T_n + \eta T_1 e^{(\pi-\lambda)/\xi} & \pi/2 \leq \lambda \leq \pi \end{cases} \tag{B.5}$$

where

$$\eta = \frac{\xi}{1+\xi^2}\frac{e^{\frac{\pi}{2\xi}} + e^{\frac{3\pi}{2\xi}}}{e^{\frac{2\pi}{\xi}} - 1}. \tag{B.6}$$

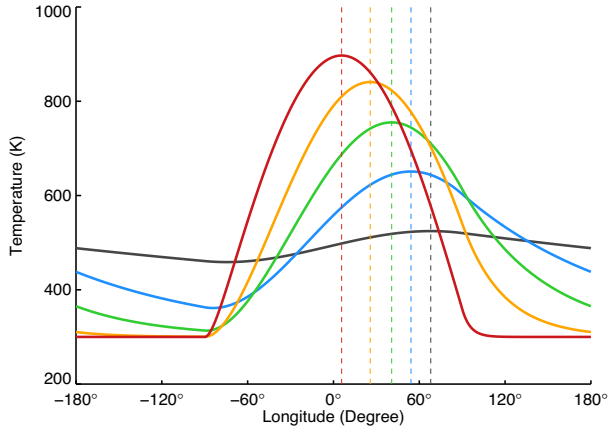

FIG. 19.— Analytical temperature solutions from Eq. (B.5) for $T_n = 300K$, $T_1 = 600K$ and different ratios between the radiative timescale and the advective timescale $\xi = \tau_{rad}/\tau_{adv}$. The red, orange, green, blue and black colors corresponds to $\xi = 0.1, 0.5, 1, 2,$ and 10, respectively. The vertical dashed lines indicate the longitudes of the temperature maximum, i.e., the"hot spot".

TABLE 3
The parameters for the $H_2$ and $CO_2$ cases in Experiment I at 700 Pa. The first three are used in the analytical calculation.

| Composition | $T_n$ (K) | $\Delta T_{eq}$ (K) | $\xi = \tau_{rad}/\tau_{adv}$ | $\tau_{rad}$ (s) | $\overline{u}$ (m s$^{-1}$) | $\tau_{adv} = R_p/\overline{u}$ (s) |
|---|---|---|---|---|---|---|
| $H_2$ | 294.7 | 471.8 | 5.1 | 43756.6 | 1999.1 | 8553.9 |
| $CO_2$ | 294.7 | 471.8 | 0.03 | 3043.4 | 176.6 | 96845.8 |

The analytical temperature distribution is not a purely sinusoidal function (Fig. 19). The temperature peak ("hot spot") on the dayside shifts relative to the equilibrium temperature pattern. When the zonal wind is eastward, $\tau_{adv}$ is positive, leading to an eastward shifted hot spot in the downwind direction. The shape of temperature curve is significantly controlled by $\xi = \tau_{rad}/\tau_{adv}$ (Fig. 19). The longitudinal shift of the hot spot becomes larger as $\xi$ increases. The day-night temperature difference is roughly reduced by a factor of $\cos^{-1}\lambda_s$ or $\sqrt{1 + \xi^2}$ compared with the radiative equilibrium situation. As $\xi$ increases, the dayside temperature decreases and nightside temperature increases, resulting a smaller day-night temperature difference. Although both this theory and the theory in Appendix A can be used to explain the temperature difference between the dayside and nightside, they are very different. The theory in Appendix A is a fully predictive theory from the first principle. But the theory we introduced here requires information of zonal-mean zonal wind $\overline{u}$, the theory of which has not been developed yet.

We can see two extreme cases where $\xi$ approaches either of the two limits: in the strong radiation regime where $\xi \sim 0$ or in the weak radiative regime where $\xi \to \infty$. If the radiative timescale is very short relative to the advective timescale, i.e., $\xi \to 0$, the temperature should be strongly controlled by the radiative relaxation. One can show that $\eta \to 0$ and $\lambda_s \to 0$ and the solution retains the equilibrium temperature distribution $T(\lambda) = T_n + T_1 \cos\lambda$ in the dayside and $T(\lambda) = T_n$ in the nightside. On the other hand, if the radiative timescale is very long compared with the advective timescale, i.e., $\xi \to \infty$, the advection will efficiently distribute the heat across longitudes. One can show that $\eta \to 1/\pi$ and $\lambda_s \to \pi/2$. In this limit, the temperature becomes constant with longitude and $T(\lambda) = T_n + T_1/\pi$.

It is straightforward to generalize the solution for a 3D atmosphere with the 3D equilibrium temperature distribution in Eq. (9). The solution is:

$$T(\lambda, \phi, p) = \begin{cases} T_n(p) + \Delta T_{eq}(p)\cos\phi\cos\lambda_s\cos(\lambda - \lambda_s) + \eta\Delta T_{eq}(p)\cos\phi\, e^{-\lambda/\xi} & -\pi/2 \leq \lambda \leq \pi/2 \\ T_n(p) + \eta\Delta T_{eq}(p)\cos\phi\, e^{-(\pi+\lambda)/\xi} & -\pi \leq \lambda \leq -\pi/2 \\ T_n(p) + \eta\Delta T_{eq}(p)\cos\phi\, e^{(\pi-\lambda)/\xi} & \pi/2 \leq \lambda \leq \pi \end{cases} \quad (B.7)$$

where $\eta$ is given by Eq. (B.6). To compare with our 3D numerical results, we adopted $T_n, \Delta T_{eq}$ and $\tau_{rad}$ the same as that in the Experiment I cases (Section 3). However, the advection timescale is assumed as $\tau_{adv} \sim R/\overline{u}$ where $R$ is the planetary radius and $\overline{u}$ is the zonal-mean zonal wind. Since a first-principle analytical theory of $\overline{u}$ has not yet been developed, we used the equatorial-mean $\overline{u}$ from the numerical results from Experiment I. With those parameters (Table 3), the analytical temperature distribution agrees well with the GCM results (Fig. 20).

The analytical thermal phase curves can be derived based on analytical thermal phase curve based on the 2D temperature solution (Eq. B.7) and the procedure described in Section 4.2 (Eqs. 17 and 18). The analytical phase curves agree well with that derived from 3D simulations (Fig. 21). The phase curve amplitudes from the analytical solution are slightly larger than the numerical results but the overall curve shapes, especially the phase shifts, match pretty well. Our theory explains the numerical results better when $\xi$ is smaller, as in the $CO_2$ case. This physically based, analytical 2D temperature solution captures key features for calculating the thermal phase curves, which can be characterized by three parameters in the atmosphere: $T_n, \Delta T_{eq}$, and the ratio $\xi$ between $\tau_{rad}$ and $\tau_{adv}$. Eq. (B.7)

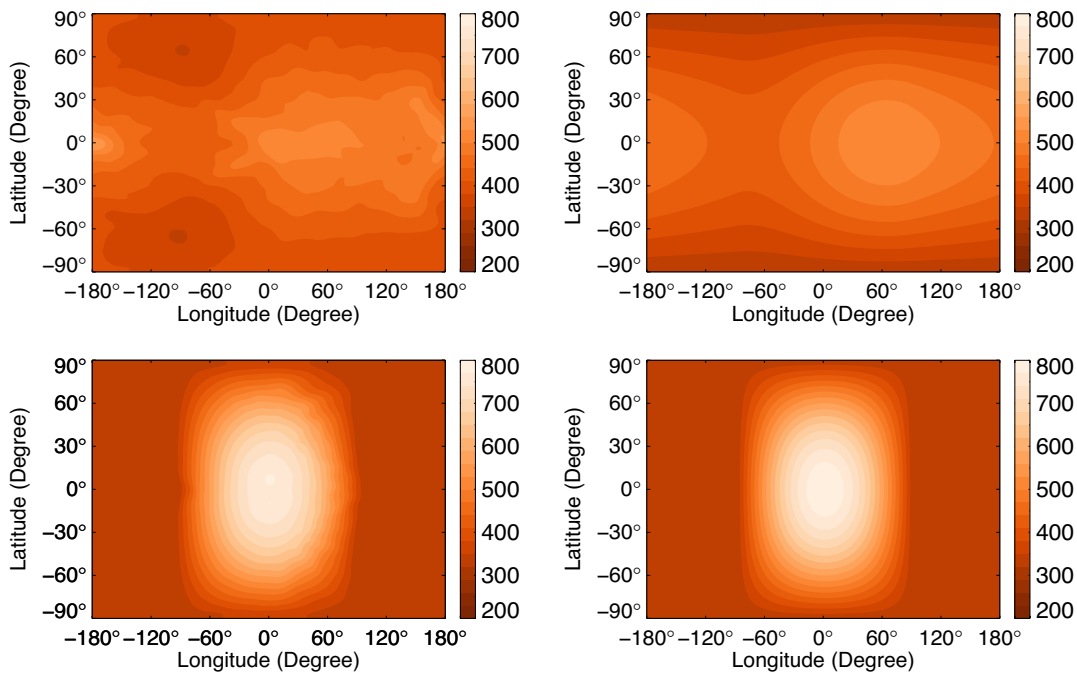

FIG. 20.— Longitude-latitude temperature (in Kelvin) maps at 700 Pa from Experiment I (left column) and the corresponding analytical temperature maps (right column) from Eq. (B.7). The top panel is for $H_2$ and bottom for $CO_2$. The advection timescale $\tau_{adv}$ used in Eq. (B.7) is derived from the simulations. The parameters used in the analytical calculation are in Table 3.

might be useful in atmospheric phase curve analysis for synchronously rotating exoplanets.

### B.2. Hot Spot Offset and Thermal Phase Shift

By setting the derivative of temperature $dT(\lambda)/d\lambda = 0$ in Eq. (B.5), one can derive the longitude of the local temperature maximum $\lambda_m$ on the dayside. This leads to the condition:

$$\sin(\lambda_s - \lambda_m)e^{\lambda_m/\xi} = \frac{\eta}{\xi\cos\lambda_s}. \tag{B.8}$$

The analytical solution of $\lambda_m$ cannot be shown in an explicit form, but it depends solely on $\xi$. As $\xi$ increases, the hot spot shifts more to the east (Fig. 19). In an eastward flow, the hot spot location is eastward of the substellar point, but by a smaller amount than the value $\lambda_s = \tan^{-1}\xi$. In other words, the hot spot location is westward of $\lambda_s$. When the hot spot shift is small ($\lambda_m$ is small), $e^{\lambda_m/\xi}$ is close to unity and $\lambda_m \approx \lambda_s = \tan^{-1}\xi$ is generally a good approximation of the hot spot location. This situation occurs when the radiative timescale is smaller than the advective timescale ($\xi$ small). When the advective timescale is smaller than the radiative timescale ($\xi$ large), the hot spot location predicted from Eq. (B.7) could be 10 to 20 degrees west of the longitude $\lambda_s$.

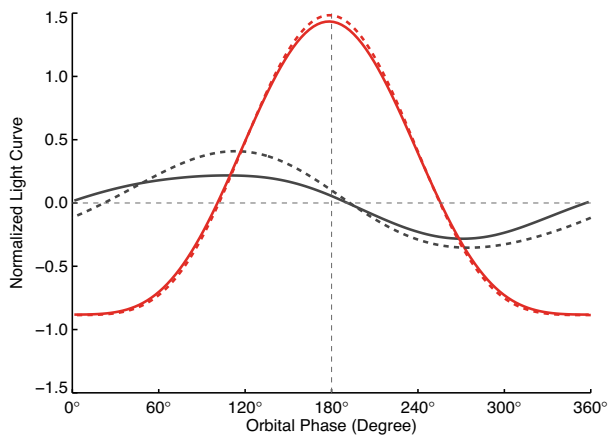

FIG. 21.— Normalized thermal phase curves of $H_2$ (black) and $CO_2$ (red) for Experiment I at 700 Pa based on the temperature distributions in Fig. (20). The solid curves are derived from the 3D simulation and dashed curves are calculated based on the analytical solution from Eq. (B. 7). The parameters used in the analytical calculation are in Table 3. The primary transit occurs at phase $0°$ and the secondary eclipse occurs at phase $180°$.

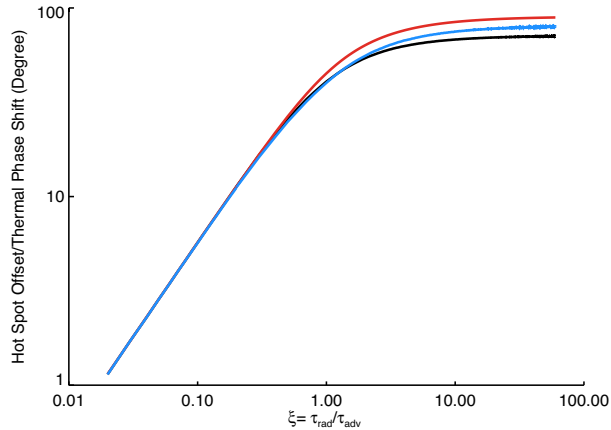

FIG. 22.— Longitudinal hot spot offset (black and red) in the temperature distribution and orbital phase shift (blue) in analytical thermal lightcurve as a function of the ratio of radiative timescale versus advection timescale. The degree of the orbital phase shift of the lightcurve was calculated based on Eqs. (17), (18), and (B.7). For the degree of the hot spot offset, the black ($\lambda_m$) is computed from Eq. (B.8) and the red is from $\lambda_s = \tan^{-1}(\tau_{rad}/\tau_{adv})$.

The analytical thermal phase curves are derived based on the 2D temperature solution (Eq. B.7) and the procedure described in Section 4.2 (Eqs. 17 and 18). We found a relationship between the longitudinal hot spot offset $\lambda_m$ and the thermal phase shift ($\delta_s$) before the secondary eclipse. When expressed in degrees, both $\lambda_m$ and $\delta_s$ increase with $\xi$ (Fig. 22) in a similar trend. When the phase shift is less than about 55 degrees, $\lambda_m$ from Eq. (B. 8) agrees perfectly with the phase shift and it might under-predict $\delta_s$ to up to 10 degrees when the phase shift is large. On the other hand, the simpler predictor $\lambda_s = \tan^{-1} \xi$ generally over-predicts $\delta_s$ when the phase shift is larger than 30 degrees. Although there are three parameters that shape the thermal lightcurve, we found that the thermal phase shift is almost totally controlled by $\xi$, whereas $T_n$ and $\Delta T_{eq}$ have negligible influence in the parameter regime we explored in this study.

### ACKNOWLEDGEMENTS

This research was partially supported by Bisgrove Scholar Program to X.Z. and NSF grant AST1313444 to A.P.S. We thank T. Kataria, T. Komacek and X. Tan for helpful discussions. Some of the simulations were performed on the Stampede supercomputer at TACC through an allocation by XSEDE.